\documentclass[aps,twocolumn,superscriptaddress,citeautoscript,nofootinbib,longbibliography]{revtex4-1}
\usepackage{graphicx}
\usepackage{multirow}
\usepackage{xcolor}
\usepackage{times}
\usepackage{amsmath,amsfonts}
\usepackage{bm}
\usepackage{bibunits}
\usepackage{tikz}
\usepackage{dsfont}
\usepackage{mathtools}
\usepackage{setspace}
\usepackage[normalem]{ulem}
\usepackage[caption=false]{subfig}

\usepackage[colorlinks=true,linkcolor=magenta,citecolor=magenta]{hyperref}

\def\ba#1\ea{\begin{align}#1\end{align}}
\def\bg#1\eg{\begin{gather}#1\end{gather}}
\def\bpm{\begin{pmatrix}}
\def\epm{\end{pmatrix}}
\newcommand{\nn}{\nonumber}
\newcommand{\tbf}[1]{\textbf{#1}}

\newcommand{\bb}[1]{{\boldsymbol #1}}
\newcommand{\bx}{{\bb x}}
\newcommand{\br}{{\bb r}}
\newcommand{\bp}{{\bb p}}
\newcommand{\bk}{{\bb k}}
\newcommand{\bu}{{\bb u}}

\newcommand{\bep}{\bb \epsilon}

\newcommand{\mc}[1]{\mathcal{#1}}
\newcommand{\mf}[1]{\mathfrak{#1}}
\newcommand{\der}{\partial}
\newcommand{\dg}{\dagger}
\newcommand{\om}{\omega}
\newcommand{\sg}{\sigma}

\newcommand{\ep}{\epsilon}

\newcommand{\ket}[1]{| #1 \rangle}
\newcommand{\bra}[1]{\langle #1 |}

\newcommand{\ra}{\rightarrow}

\newcommand{\Rf}[1]{Ref.~\onlinecite{#1}}
\newcommand{\Sec}[1]{Sec.~\ref{#1}}
\newcommand{\eq}[1]{Eq.~\eqref{#1}}
\newcommand{\eqs}[1]{Eqs.~\eqref{#1}}
\newcommand{\fig}[1]{Fig.~\ref{#1}}

\newcommand{\magenta}[1]{\textcolor{magenta}{#1}}
\newcommand{\ourtitle}{Topological acoustic triple point}

\begin{document}
\title{\ourtitle}
\author{Sungjoon \surname{Park}}
\thanks {These authors contributed equally to this work.}
\affiliation{Center for Correlated Electron Systems, Institute for Basic Science, Seoul 08826, Korea}
\affiliation{Department of Physics and Astronomy, Seoul National University, Seoul 08826, Korea}
\affiliation{Center for Theoretical Physics (CTP), Seoul National University, Seoul 08826, Korea}

\author{Yoonseok \surname{Hwang}}
\thanks {These authors contributed equally to this work.}
\affiliation{Center for Correlated Electron Systems, Institute for Basic Science, Seoul 08826, Korea}
\affiliation{Department of Physics and Astronomy, Seoul National University, Seoul 08826, Korea}
\affiliation{Center for Theoretical Physics (CTP), Seoul National University, Seoul 08826, Korea}

\author{Hong Chul \surname{Choi}}
\affiliation{Center for Correlated Electron Systems, Institute for Basic Science, Seoul 08826, Korea}
\affiliation{Department of Physics and Astronomy, Seoul National University, Seoul 08826, Korea}

\author{Bohm-Jung Yang}
\email[Electronic address:$~~$]{bjyang@snu.ac.kr}
\affiliation{Center for Correlated Electron Systems, Institute for Basic Science, Seoul 08826, Korea}
\affiliation{Department of Physics and Astronomy, Seoul National University, Seoul 08826, Korea}
\affiliation{Center for Theoretical Physics (CTP), Seoul National University, Seoul 08826, Korea}

\date{\today}
\let\oldaddcontentsline\addcontentsline
\renewcommand{\addcontentsline}[3]{}

\begin{abstract}
Acoustic phonon in a crystalline solid is a well-known and ubiquitous example of elementary excitation with a triple degeneracy in the band structure.
Because of the Nambu-Goldstone theorem, this triple degeneracy is always present in the phonon band structure.
Here, we show that the triple degeneracy of acoustic phonons can be characterized by a topological charge $\mf{q}$ that is a property of three-band systems with $\mc{PT}$ symmetry, where $\mc{P}$ and $\mc{T}$ are the inversion and the time-reversal symmetries, respectively.
We therefore call triple points with nontrivial $\mf{q}$ the topological acoustic triple point (TATP).
The topological charge $\mf{q}$ can equivalently be characterized by the skyrmion number of the longitudinal mode, or by the Euler number of the transverse modes, and this strongly constrains the nodal structure around the TATP.
The TATP can also be symmetry-protected at high-symmetry momenta in the band structure of phonons and spinless electrons by the $O_h$ and the $T_h$ groups.
The nontrivial wavefunction texture around the TATP can induce anomalous thermal transport in phononic systems and orbital Hall effect in electronic systems. 
Our theory demonstrates that the gapless points associated with the Nambu-Goldstone theorem are an avenue for discovering new classes of degeneracy points with distinct topological characteristics.
\end{abstract}

\maketitle

Classification of topological phases of matters has been a topic of intensive research~\cite{hasan2010colloquium,po2017symmetry,bradlyn2017topological}.
An important conclusion drawn from these investigations is that gap closing points in the band structure are often characterized  by a topological charge. 
A famous example is the Weyl point, whose gaplessness is protected by the Chern number \cite{armitage2018weyl}.
However, in nature, there is a different class of gap closing points that are enforced by the Nambu-Goldstone (NG) theorem, whose topological characteristics have been largely unexplored yet.
\cite{nambu1961dynamical,goldstone1961field,goldstone1962broken}.
A familiar example is the acoustic phonons, which are NG bosons resulting from breaking the translational symmetries, and exist even in classical systems.
Because three translational symmetries are broken, there are three gapless acoustic phonons forming a triple point at the Brillouin zone (BZ) center, which we refer to as the acoustic triple point (ATP).

Here, we show that an ATP can carry a topological charge $\mf{q}$, which consists of a pair of well-known topological charges: the skyrmion number $\mf{n}_{sk}$ and the Euler number $\mf{e}$. 
Hence, an ATP with nontrivial $\mf{q}$ is dubbed the `topological ATP' (TATP).
The topological charge $\mf{q}$ is strictly defined only when the total number of energy bands is fixed to three, so that it falls under the recently proposed category of `delicate' (topological) charge~\cite{nelson2020multicellularity}, which is distinct from the stable charge~\cite{kitaev2009periodic,chiu2016classification} such as the Chern number or the fragile charge~\cite{po2018fragile,liu2019shift,bradlyn2019disconnected,bouhon2019wilson,hwang2019fragile,song2020fragile,song2020twisted} such as the Euler number~\cite{ahn2019failure}.
In general, the delicate charge is defined for small number of bands and thus it is not well-defined in electronic system, where the total number of electron bands easily exceeds the relevant number.
In contrast, the number of phonon energy bands is fixed by the number of atoms in the unit cell, so that there is a possibility that phonons can be exactly characterized by the delicate charge.

The TATP protected by NG theorem exists ubiquitously in elastic material.
Interestingly, the triple points with nontrivial $\mf{q}$ can also be symmetry-protected at high symmetry momentum in $\mc{PT}$ symmetric elastic systems, and even in $\mc{PT}$ symmetric electronic systems with negligible spin-orbit coupling,  where $\mc{P}$ and $\mc{T}$ are inversion and time-reversal symmetries, respectively.
The TATP protected by the NG theorem has a linear dispersion, while the symmetry-protected triple point has a quadratic dispersion around the triple point. 
However, since both share the same topological charge, we refer to both types of triple points as the TATP.

\begin{figure}[t]
\centering
\includegraphics[width=8.5cm]{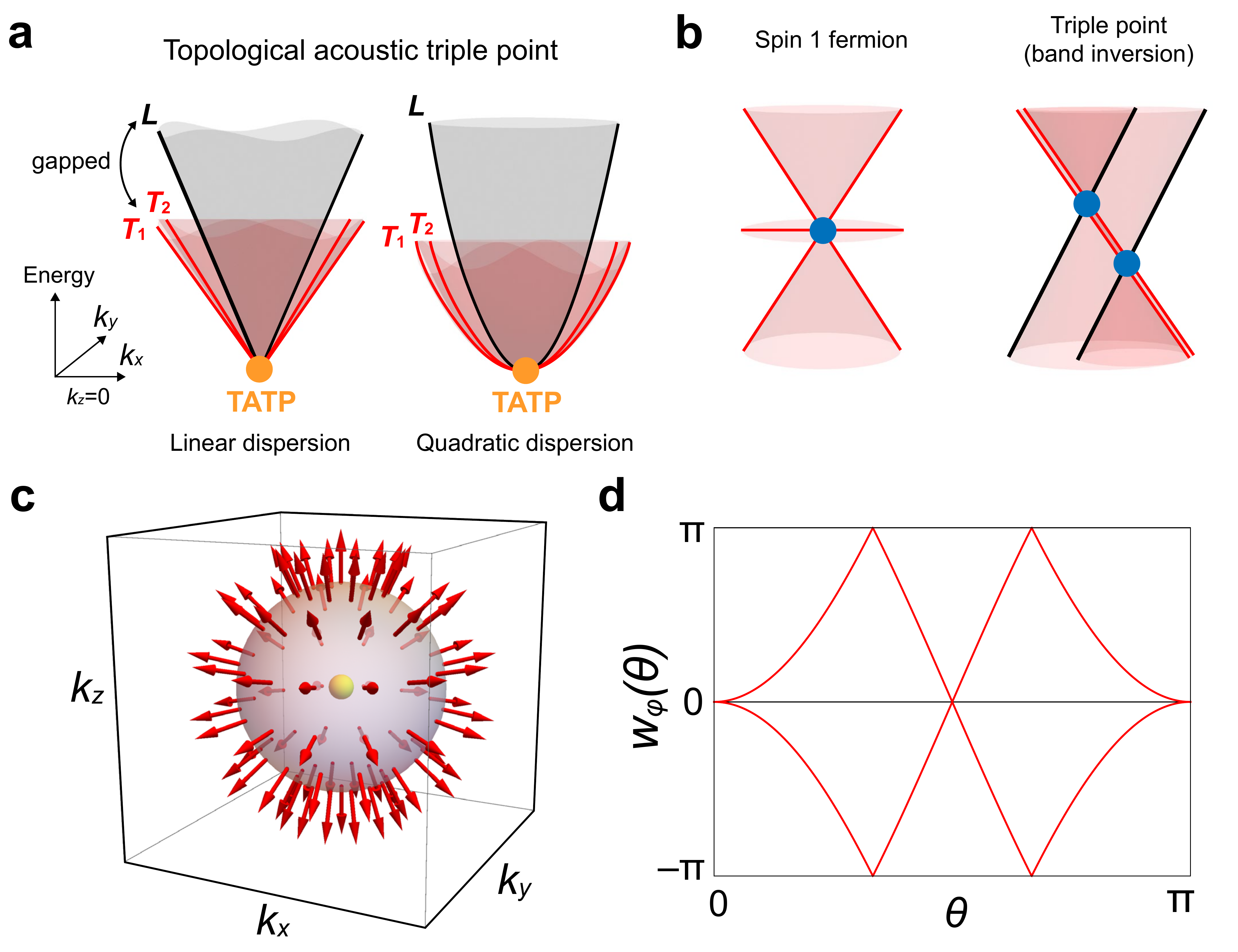}
\caption{\tbf{Topological acoustic triple point (ATP).} 
\tbf{a,} Topological ATP (TATP) with $\mf{q}\equiv (\mf{n}_{sk},\mf{e})=(1,2)$ can appear with either a linear dispersion or a quadratic dispersion. 
Note that there is a gap between the $L$ and the $T$ modes away from the triply degenerate point.
\tbf{b,} TATP is distinct from spin-1 Weyl point, which is protected by Chern numbers.
TATP is also distinct from the triple point formed by band inversion, where it is not possible to separate the highest energy band from two lower energy bands away from the triple point.
\tbf{c,} Skyrmion texture of longitudinal phonon on the sphere wrapping a TATP. The transverse modes span the tangent plane of the sphere.
\tbf{d,} Wilson loop spectrum for the transverse modes as a function of the polar angle $\theta$ computed on a sphere wrapping the TATP.
$|\mf{e}|$ is given by the number of times that one of the two branches of $w_\phi(\theta)$ crosses $\pi$.
}
\label{Fig_1}
\end{figure}

A characteristic feature of both the linearly and quadratically dispersing TATPs is the  energy gap between the highest energy band ($L$ mode) and the two lower energy bands ($T$ modes), except at the triple point, see \fig{Fig_1}a.
This gap is necessary to define the topological charge $\mf{q}$, and this feature distinguishes the TATPs from the triple points created by band inversion ~\cite{zhu2016triple,wang2017prediction,lv2017observation,ma2018three,kim2018nearly,winkler2019topology,das2020topological,lenggenhager2020multi} and the spin-1 Weyl point \cite{bradlyn2016beyond,chang2017unconventional,tang2017multiple,rao2019observation,rao2019observation,rao2019observation,takane2019observation,miao2018observation}, see \fig{Fig_1}b.

Because having a nontrivial $\mf{q}$ implies nontrivial $\mf{n}_{sk}$ and $\mf{e}$ for the longitudinal and the transverse modes, respectively, it has interesting consequences for the nodal structure.
For example, there must be at least four nodal lines formed between the $T$ modes emanating from the TATP.
Also, because the nonzero $\mf{q}$ is accompanied by nontrivial winding texture of the wavefunctions around the TATPs, systems with TATPs can show anomalous transport of phonon angular momentum or electronic orbital.

\section{Topological charge}
Let $H_\bk$ denote either the dynamical matrix of phonon or the Hamiltonian matrix of electron, and let $\epsilon_{\bm{k},n}$ and $\mc{E}_{\bm{k},n}$ be the eigenvector and eigenvalue of $H_\bk$, respectively. 
Note that when $H_\bk$ is the dynamical matrix of phonon, $\mc{E}_{\bm{k},n}=\omega_{\bm{k},n}^2$ where $\omega_{\bm{k},n}$ is the phonon energy, and when  $H_\bk$  is the Hamiltonian matrix of electron, $\mc{E}_{\bm{k},n}$ is the electron energy.
This means that insofar as the topology of ATP is concerned, there is no difference between the dynamical matrix of phonon and the Hamiltonian matrix of electron.
Henceforward, we blur the difference between phonon and electron and refer to $H_\bk$ as the Hamiltonian, and clarify the difference when a possibility of confusion arises.

To define $\mf{q}$, we assume that $H_\bk$ is a $3 \times 3$ real symmetric matrix.
This condition is strictly satisfied by the phonons in a monatomic lattice, which have only three phonon bands.
Even when there is more than one atom per unit cell, and therefore more than three phonon energy bands, this condition is satisfied near the ATP, which can be described by the elastic continuum Hamiltonian \cite{lifshitz1986theory}.
Since the elastic continuum Hamiltonian is always real symmetric, independently of the crystalline symmetry, $\mf{q}$ can be defined for acoustic phonons in any elastic material.
When the TATP appears as a symmetry-protected degeneracy, $\mc{PT}$ symmetry is necessary for both elastic systems and electronic systems with negligible spin-orbit coupling.

For concreteness, let $H_\bk$ take the form of the dynamical matrix of isotropic elastic continuum,
\ba
[H_\bk]_{\alpha \beta} = v_T^2 k^2 \delta_{\alpha \beta} + (v_L^2 - v_T^2)k_\alpha k_\beta, \label{eq.iso_H}
\ea
where $v_L$ and $v_T$ are the longitudinal and transverse velocities, respectively. 
Defining $k=\sqrt{k_x^2+k_y^2+k_z^2}$ and $\tilde{k} = \sqrt{k_x^2+k_y^2}$, the eigenstates are $\bep_{\bk,L}= \frac{1}{k} (k_x,k_y,k_z)$, 
$\bep_{\bk,T_1} = \frac{1}{\tilde{k}}(-k_y,k_x,0)$,
$\bep_{\bk,T_2} =\frac{1}{\tilde{k}k}(-k_x k_z,-k_y k_z,k_x^2+k_y^2)$, whose eigenvalues are given by $v_L^2 k^2$, $v_T^2 k^2$, and $v_T^2 k^2$ respectively.
To define $\mf{q}$, we consider a sphere surrounding the triple point at $k=0$.
On this sphere, notice that the  $L$ mode has the skyrmion number $\mf{n}_{sk}=1$, see \fig{Fig_1}c.
Furthermore, the $T$ modes span the tangent space of the sphere in the momentum space, so that they have Euler number $\mf{e}=2$ as is well-known.
Alternatively, the Euler number can be computed by counting the winding number of the Wilson loop spectrum \cite{bzduvsek2017robust,ahn2018band}, see \fig{Fig_1}d . 
Therefore, we define the topological charge $\mf{q}=(\mf{n}_{sk},\mf{e})$, where it can be shown that the constraint $\mf{e}=2\mf{n}_{sk}$ must be satisfied.
This discussion can be generalized to any $3 \times 3$ real symmetric Hamiltonian as long as there is a gap between the $L$ and the $T$ modes, see Methods and Supplementary Information (SI)~\cite{supplement}.

\begin{figure}[t]
\centering
\includegraphics[width=8.5cm]{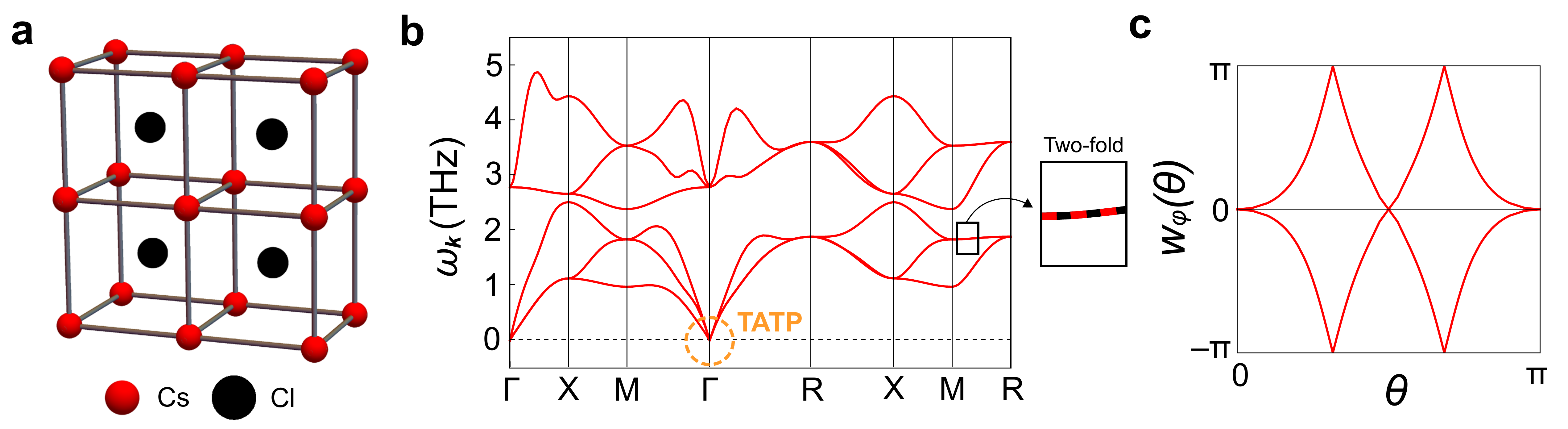}
\caption{\tbf{TATP in CsCl.} \tbf{a,} CsCl lattice structure.
\tbf{b,} Phonon spectrum ($\omega_{\bm{k}}$) of CsCl along the high symmetry lines obtained by first-principles calculations.
Acoustic phonons at $\Gamma$ carry $\mf{q}=(1,2)$.
$\mf{q}$ cannot be defined for the triple degeneracy at the $R$ point, because lower two bands cannot be fully separated from the highest energy band 
owing to the degeneracy between upper two bands along the $RM$ direction.
\tbf{c,} Wilson loop spectrum of the two lowest acoustic phonons near the $\Gamma$ point.
}
\label{Fig_2}
\end{figure}

\section{Phonons in C\lowercase{s}C\lowercase{l}}
Although the topological charge $\mf{q}$ was defined for a three-band system, the topological charge is still meaningful in multiband systems.
To demonstrate this, let us study the phonon spectrum of CsCl lattice, which has two atoms per unit cell (see \fig{Fig_2}a), and therefore six phonon bands.
We show the phonon spectrum ($\omega_{\bm{k}}$) obtained from first-principles calculations (see Methods) in \fig{Fig_2}b.
Near $\Gamma$ with $\bm{k} \neq 0$, we see that the two lowest acoustic modes are gapped from the others.
Therefore, we can compute the Wilson loop spectrum for these two acoustic phonons, which we show in \fig{Fig_2}c.
From the winding structure of the Wilson loop spectrum, we see that $|\mf{e}|=2$.
It is important to note that although the CsCl lattice has six phonon bands, we can still define the Euler number for the $T$ modes.
This is because $\mf{e}$ can be defined for any two bands that are isolated from the others by a gap, so that it is not sensitive to the total number of energy bands present in the system.
In contrast, $\mf{n}_{sk}$ and $\mf{q}$ are properties of a $3 \times 3$ Hamiltonian, so that they are not well-defined here in a strict sense.
Therefore, $\mf{q}=(\mf{n}_{sk},\mf{e})$ reduces to $\mf{e}$.
However, we can recover the topological charge $\mf{q}$ in the low-energy continuum limit, as we discuss in the next section.

We note that the triply degenerate optical modes at $\Gamma$ is also a TATP (see Methods), whereas $\mf{q}$ cannot be defined for the triple degeneracy at the $R$ point, because lower two bands cannot be fully separated from the highest energy band (see \fig{Fig_2}b).

\section{Continuum theory}
In this section, we discuss how the continuum theory constrained by the crystalline symmetries allows us to extend the discussion of TATP to general multi-band systems.
Let us first consider the gapless acoustic phonons, which are conventionally described by the elastic continuum theory.
This naturally yields a $3\times 3$ effective Hamiltonian (dynamical matrix) description of the acoustic phonons, whose specific form is constrained by the 32 point group symmetries allowed by the crystal \cite{nye1985physical}.
Because the triple point is always present due to the gaplessness of phonons, all 32 point group symmetries are meaningful.

\begin{figure*}[t]
\centering
\includegraphics[width=12cm]{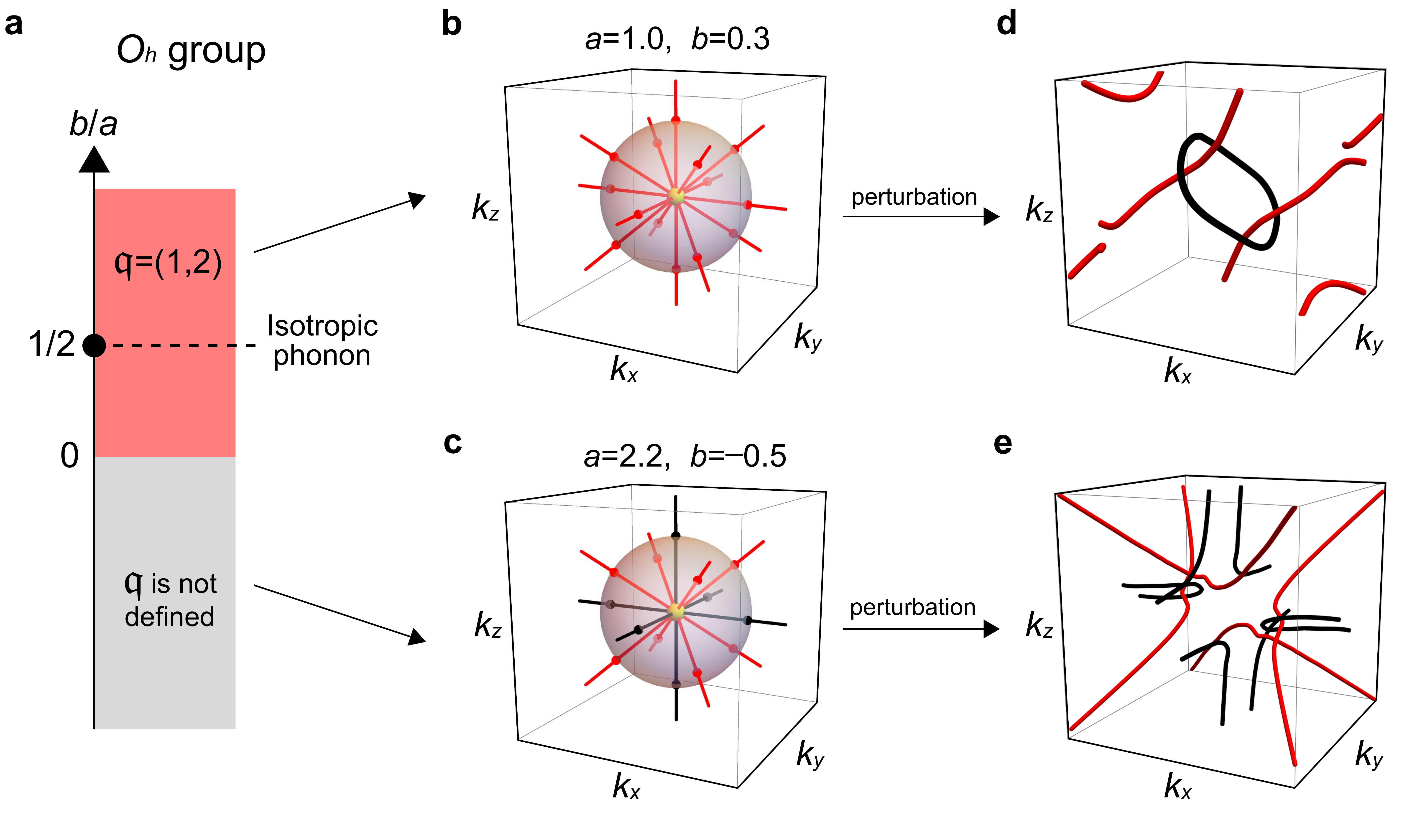}
\caption{\tbf{ATPs in cubic systems.} 
\tbf{a,} Phase diagram for the elastic continuum Hamiltonian in \eq{eq.effective} for acoustic phonons in cubic systems.
\tbf{b, c,} Nodal structure for $b/a>0$ (\tbf{b}) and $b/a<0$ (\tbf{c}).
The black (red) lines are band degeneracies between the upper (lower) two bands.
Notice that there are two types of nodal lines, one along the $k_x$, $k_y$, and $k_z$ axes and another along the lines that satisfy $|k_x|=|k_y|=|k_z|$.
For $b/a>0$, the band degeneracies occur only between the lower two bands.
However, the eigenvalues of the degenerate bands along the $k_x$, $k_y$, and $k_z$ axes increases as $b/a$ decreases, 
so that when $b/a<0$, these degeneracies occur between the upper two bands instead of the lower two bands.
When we perturb the Hamiltonian in \tbf{b} such that the conditions required to obtain the  symmetry-protected TATP are broken, while the conditions needed to define $\mf{q}$ are kept, we obtain \tbf{d}.
Notice that the nodal ring (black) formed between the upper two bands are penetrated by two nodal lines formed between the lower two bands.
This should be compared with \tbf{e}, in which we do not obtain a linked nodal ring structure as in \tbf{d}, although we similarly perturb the Hamiltonian in \tbf{c}.
}
\label{Fig_3}
\end{figure*}

For simplicity, let us focus on the effective Hamiltonian of  the elastic continuum constrained by the cubic symmetries.
Because we are interested in the topological properties, it is sufficient to examine only the traceless part of the Hamiltonian, which takes the form
\ba
H_\bk = \sum_n f_n(\bk) \lambda_n, \label{eq.effective}
\ea
where $n = 1,3,4,6,8$ and $\lambda_n$ are the Gell-Mann matrices.
For cubic groups, we find $f_1(\bk) =a k_x k_y$, $f_3 (\bk) = b (k_x^2 - k_y^2)$, $f_4(\bk) = a k_x k_z$, $f_6(\bk)= a k_y k_z$, $f_8(\bk)= \tfrac{b}{\sqrt{3}}(k_x^2 + k_y^2) - \tfrac{2b}{\sqrt{3}} k_z^2$.
Here, $a$ and $b$ are constants that can be related to the three elastic constants of a cubic crystal, $C_{11}$, $C_{12}$ and $C_{44}$, by the following relations: $a=C_{12}+C_{44}$ and $b=\tfrac{C_{11}}{2}-\tfrac{C_{44}}{2}$.
When $a \neq 0$, the topological properties of the Hamiltonian are determined by only one parameter, $b/a$, so that we can draw a phase diagram as shown in \fig{Fig_3}a.
We find that $b/a>0$ corresponds to the band structure shown in \fig{Fig_1}a, so that the $L$ mode is gapped from the $T$ modes for $k>0$, and the topological charge is $\mf{q}=(1,2)$.
When $b/a<0$, $\mf{q}$ is not defined because it is no longer possible to properly partition the energy bands for $k>0$, see \fig{Fig_3}b,c.

The criterion $b/a>0$ allows us to easily search for materials with TATP.
In particular, the acoustic phonons of monatomic lattices such as Au, Ag, and Cu are topological with $\mf{q}=(1,2)$.
Since monatomic lattices have a total of three phonon modes, the topological charge $\mf{q}$ in these materials can be defined without using the continuum approximation.

It turns out that the above condition that the phonons carry $\mf{q}$ amounts to the condition that the longitudinal velocity exceeds the transverse velocity along the high symmetry lines~\cite{supplement}.
For isotropic systems, the transverse velocity cannot exceed the longitudinal velocity because of the Born stability condition for isotropic systems that $v^{2}_{T}/v^{2}_{L}<3/4$.
However, the Born stability criteria of cubic crystals~\cite{born1940stability,born1954dynamical} do not forbid  $v_T>v_L$ along the high symmetry lines 
so that it is possible to observe acoustic phonons which do not carry $\mf{q}$. 
The necessary and sufficient conditions for stability of cubic crystals are~\cite{mouhat2014necessary} $C_{44}>0$, $C_{11}-C_{12}>0$, $C_{11}+2C_{12}>0$, which allows $v_T>v_L$.
Indeed, such situations are known to occur~\cite{every1985phonon} in certain Tm-Se and Sm-Y-S intermediate valence compounds~\cite{boppart1980first,mook1979neutron} and certain Mn-Ni-C alloys~\cite{lowde1981martensitic,sato1981martensitic}.

Although TATP can appear for any crystal symmetry for acoustic phonons, the symmetry-protected TATPs of phonons, or of electrons, require stricter symmetry constraints.
Of the 32 point group symmetries, only the $O_h$ and the $T_h$ groups contain the inversion symmetry and support three-dimensional representations.
In the case of the $O_h$ group (see the SI~\cite{supplement} for $T_h$ group), four representations ($T_{1u}$, $T_{2u}$, $T_{1g}$, $T_{2g})$ allow a triple point, and the effective Hamiltonian near the triple point takes the form in Eq.~\eqref{eq.effective} after appropriate transformations.
Therefore, the phase diagram in \fig{Fig_3} applies here as well.

\section{Nodal structure}
The charge $\mf{q}=(\mf{n}_{sk},\mf{e})$ strongly constrains the nodal structure.
As before, we consider a sphere on which $\mf{q}$ is nontrivial.
First, because the skyrmion number of the $L$ mode cannot change under a continuous deformation of the Hamiltonian without closing the gap, the $L$ mode must cross the $T$ modes inside the sphere, which occurs at the ATP.
Second, the Euler number constrains the number of nodal lines formed between the $T$ modes that pass through the ATP.
This is because nodal lines emanating from the TATP can be considered as Dirac points on the 2D sphere surrounding the TATP, and nonzero $\mf{e}$ constrains the total vorticity $N_t$ (signed count of the number of Dirac points) to be $N_t = -2 \mf{e}$  \cite{ahn2019failure}.
Thus, when $\mf{e}=2$ for the $T$ modes, $N_t=-4$, so that there must be at least four nodal lines emanating from the TATP,  see the Methods.

At this point, it is interesting to note that the presence of symmetry-protected TATPs requires more constraints than it is needed to define  $\mf{q}$.
This is because the definition of $\mf{q}$ only requires that the Hamiltonian be a $3 \times 3$ real symmetric matrix with a spectral gap between $L$ and the $T$ modes, while the symmetry-protected TATP requires further constraints such as the $O_h$ symmetry.
Thus, it is natural to ask how the topological charge $\mf{q}=(1,2)$ constrains the nodal structure of symmetry-protected TATPs when we perturb the Hamiltonian so that the relevant symmetry is relaxed, while the conditions required to define $\mf{q}$ are maintained.
In \fig{Fig_3}d, we show the nodal structure that results from adding such perturbations to the Hamiltonian used in \fig{Fig_3}b.
As explained further in the Methods, we find a nodal ring formed between the $L$ and the $T$ modes (black ring) that is threaded by two nodal lines formed between the $T$ modes (red lines).
For comparison, we similarly perturb the Hamiltonian for the case where $\mf{q}$ is ill-defined, used in \fig{Fig_3}c.
The resulting nodal structure is shown in \fig{Fig_3}e.
Although the nodal structure is complicated, we see that there is no linking structure similar to that observed in \fig{Fig_3}d.

\section{Avoiding the doubling theorem}
\begin{figure}[t]
\centering
\includegraphics[width=8.5cm]{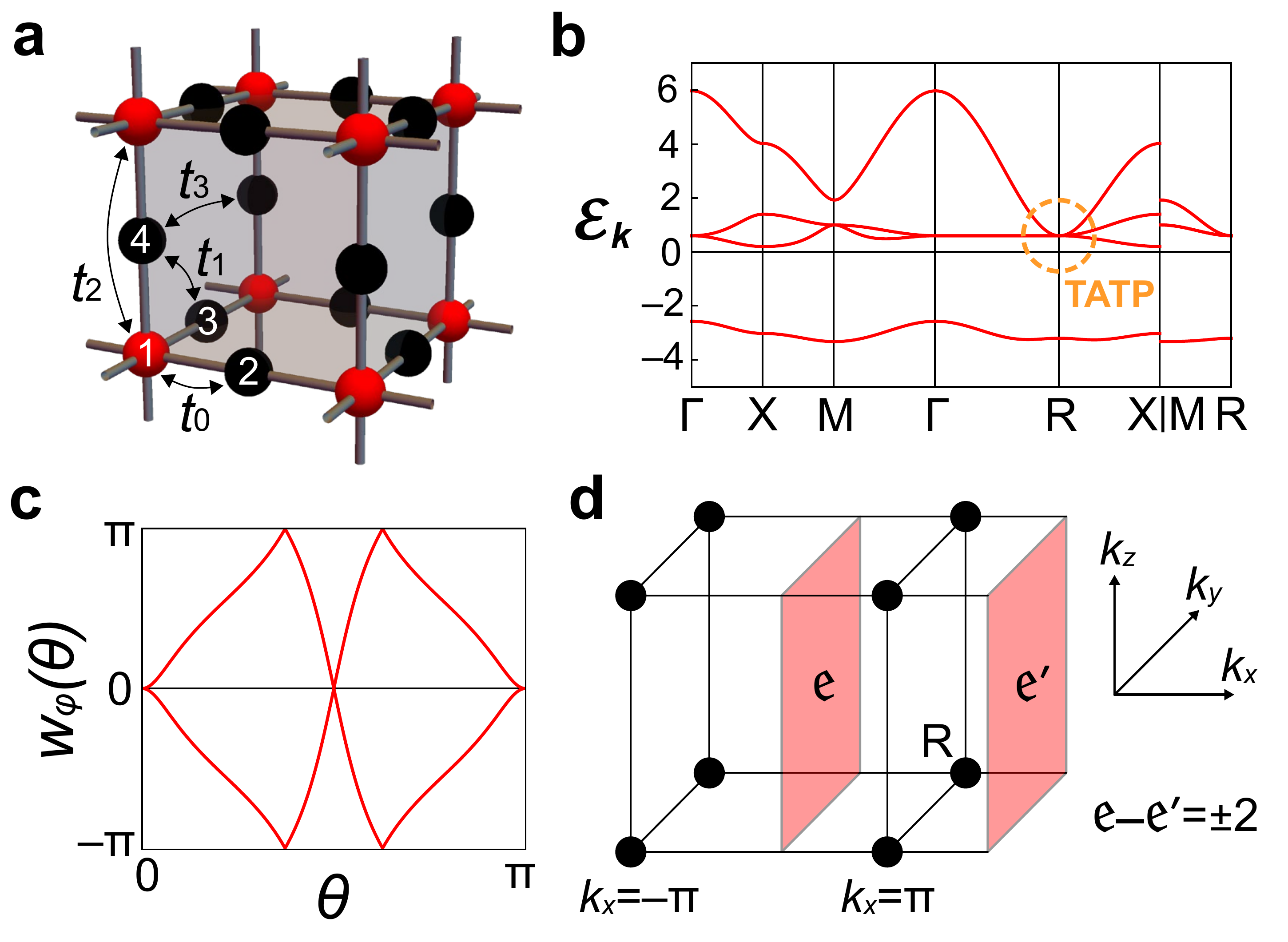}
\caption{
\tbf{Avoiding the doubling theorem in 3D electronic Lieb model.} \tbf{a,} 3D Lieb lattice structure with four sites in a unit cell. 
\tbf{b,} The electronic band structure with a single TATP at $R$. 
Here $\Gamma=(0,0,0)$, $X=(\pi,0,0)$, $M=(\pi,\pi,0)$, and $R=(\pi,\pi,\pi)$. 
We refer to the highest energy mode forming the TATP as the $L$ mode, and the lower two energy modes forming the TATP as the $T$ modes.
\tbf{c,} Wilson loop spectrum for the second and third lowest bands, corresponding to $T$ mode of the TATP, computed over a sphere with radius $0.1\pi$ centered at $R$. 
The winding structure shows that $|\mf{e}|=2$.
\tbf{d,} With a single TATP at $R$ (black dots), it is not possible to define $\mf{e}$ in the $(k_y,k_z)$ plane since it conflicts with the periodicity of the Brillouin zone.
This contradiction is resolved by noticing the $\pi$ Zak phases along the $k_x$, $k_y$, $k_z$ directions, so that $\mf{e}$ is ill-defined.
}
\label{Fig_4}
\end{figure}

It is well known that Weyl points must appear in pairs because of the Nielsen-Ninomiya theorem~\cite{nielsen1981no}, which is simply the result of the periodicity of the BZ and the topological charge (Chern number) that protects the Weyl points.
Likewise, because the TATPs are protected by the topological charge $\mf{q}$, one might expect that TATPs will appear in pairs.
However, the doubling theorem can be avoided in various ways.

For instance, the phonon spectrum of CsCl in \fig{Fig_2}b demonstrates one way to avoid the doubling of TATPs.
First, let us note that there is a gap between the lowest three phonon modes (acoustic phonons) and the rest of the phonon modes (optical phonons).
Focusing on the acoustic phonons, we see that there are two triple points at $\Gamma$ and $R$.
However, the triple point at $\Gamma$ is topological while that at $R$ is not.
This is allowed because there is a nodal line along $RM$ formed between $L$ and the $T$ modes.
Because this nodal line stretches across each of the $k_x$, $k_y$, and $k_z$ directions, it is not possible to choose a 2D plane such that there is a gap between the $L$ and the $T$ modes. 
Therefore, it is not possible to define $\mf{e}$ for the lowest two bands on any 2D plane.

Interestingly, the doubling theorem can be avoided even when there is a gap between the $L$ and the $T$ modes in the whole BZ except at a TATP. 
Since $\mf{e}$ is defined only for an orientable vector bundle, it is well-defined only when the Zak phase is trivial for the $T$ modes along any line in the BZ~\cite{ahn2019failure}.
Therefore, a single TATP can appear in the presence of $\pi$ quantized Zak phase for the $T$ modes.
We demonstrate this in the electronic spectrum of the 3D generalization of the Lieb lattice~\cite{lieb1989two,weeks2010topological}, whose lattice structure is shown in \fig{Fig_4}a.
From the resulting band structure and the Wilson loop spectrum in \fig{Fig_4}b,c, we see that there is a single TATP at $R$, although the $L$ and the $T$ modes are fully gapped except at $R$.
We numerically confirmed that there is $\pi$ quantized Zak phase for the $T$ modes along each of the  $k_x$, $k_y$, and $k_z$ direction, which allows a single TATP.

\begin{figure}[b]
\centering
\includegraphics[width=8.5cm]{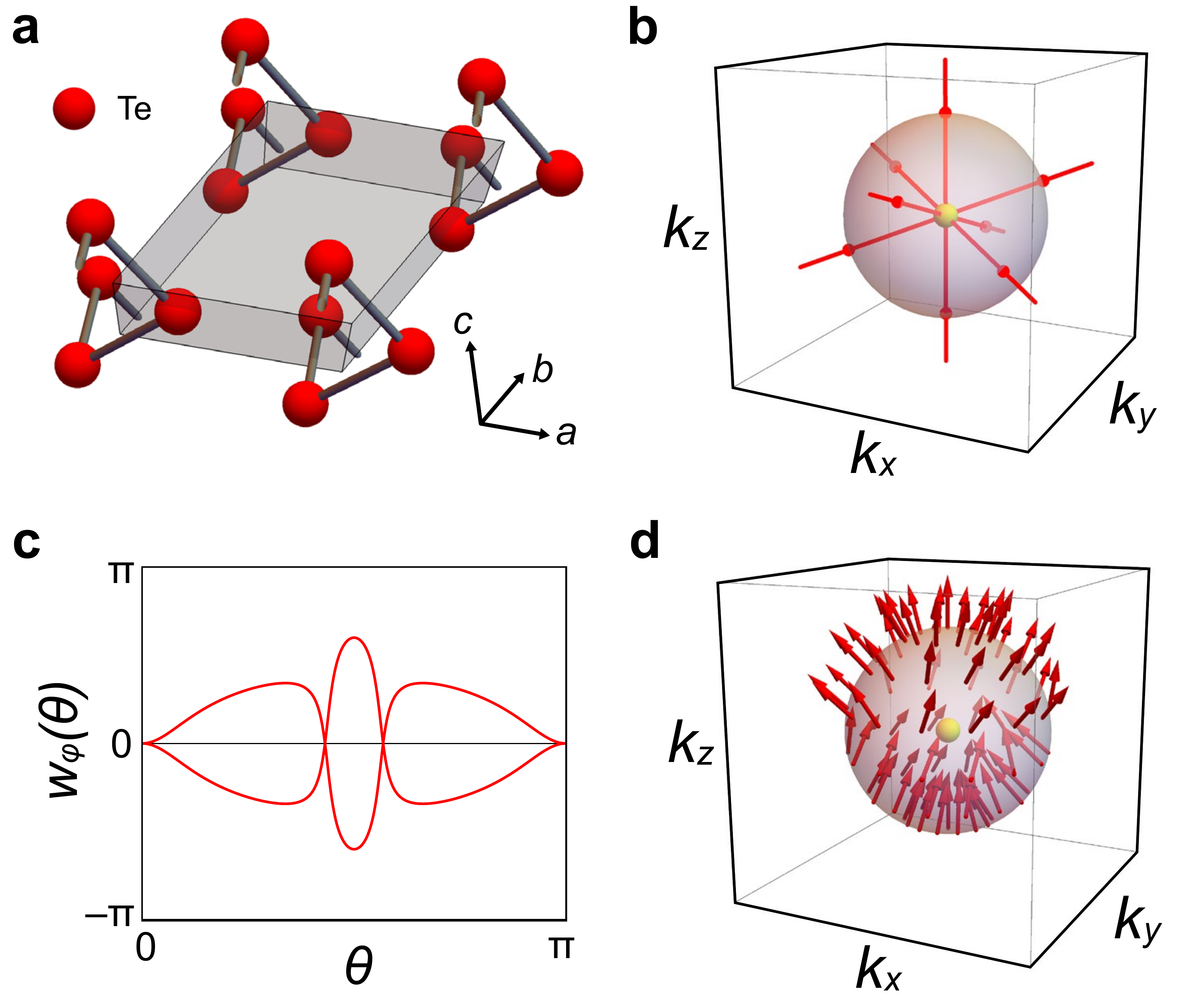}
\caption{
\tbf{Acoustic phonon of tellurium.}
\tbf{a,} The lattice structure of tellurium with space group P3${_1}$21. 
\tbf{b.} By using the values of the stiffness tensor in Materials Project \cite{de2015charting}, we find that away from the triple point, the nodal lines occur only between the two lowest energy bands so that $\mf{q}$ can be defined.
\tbf{c,} The Wilson loop spectrum shows trivial winding structure, and therefore, $\mf{e}=0$.
\tbf{d,} The wavefunction texture of the highest energy band shows trivial skyrmion texture, which is consistent with the constraint that $2\mf{n}_{sk}=\mf{e}$.
}
\label{Fig_5}
\end{figure}

\section{Discussion} 

For phonons in cubic crystals, either $\mf{q} = (1,2)$ or $\mf{q}$ is undefined.
However, when the symmetry of the crystal is sufficiently low, it is also possible to obtain $\mf{q}=(0,0)$.
The acoustic phonon of tellurium with space group  P3${_1}$21 is one such example\footnote{Tellurium lacks the inversion symmetry, so that strictly speaking, $\mf{q}$ is not defined. 
However, in the elastic continuum approximation, the inversion symmetry is restored, and this does not nullify our discussion.}, as we illustrate in \fig{Fig_5} (see the SI~\cite{supplement} for the details).
We show the gap closing points in the acoustic phonon spectrum in \fig{Fig_5}b.
Notice that the gap closing points occur only between the $T$ modes, so that we can define $\mf{q}$.
From the winding structure of the Wilson loop spectrum shown in \fig{Fig_5}c, we see that $\mf{e}=0$, which is consistent with the trivial skyrmion texture of $\mathbf{\epsilon}_{\bm{k},L}$, see \fig{Fig_5}d.

Because topological charge is often associated with surface states, it is natural to ask whether there are relevant surface states.
Since surface acoustic wave is well-known surface states related to acoustic phonons, one may suspect that it is related to $\mf{q}$.
For an isotropic medium, the stability of the material imposes the condition $v_T^2/v_L^2< 3/4$, while the condition for the appearance of surface acoustic waves is $v_T^2/v_L^2<1$.
Therefore, isotropic elastic materials satisfying the stability condition always have surface acoustic waves~\cite{lifshitz1986theory}.
However, because isotropic phonon is topological even for $v_T^2/v_L^2>1$, topology does not seem to be directly related to the surface localized states.
To further confirm this, we study the finite size 3D Lieb lattice model.
As we discuss in detail in the SI~\cite{supplement}, we find that even when the parameters are chosen so that the continuum theory for the TATP at $R$ becomes the same as the continuum theory for the isotropic phonon, there are no surface localized states.
Because the same continuum theory does not lead to the same boundary states, we conclude that surface acoustic waves result from the boundary condition specific to elastic systems.

Although nontrivial $\mf{q}$ is not directly related to surface states, it can induce anomalous transport phenomena, such as the phonon angular momentum Hall effect.
As shown in \Rf{park2020phonon}, the winding structure of isotropic phonon has a characteristic phonon angular momentum Hall response.
As a consequence, there is an edge accumulation of phonon angular momentum, which can have significant contributions from both the bulk and surface localized phonons, see the SI~\cite{supplement}.
Because the phonon angular momentum Hall effect and the orbital Hall effects are analogous, the TATPs consisting of $p$ or $d$ electron orbitals are also a significant source of the orbital Hall effect \cite{go2018intrinsic,jo2018gigantic}.
Further investigating the physical consequences of having different topological characterizations of TATPs will be an interesting topic for future study.

\vspace{1cm}
\tbf{Acknowledgements}
S.P. thanks Sunje Kim for useful discussion.
S.P., Y. H. and B.-J.Y. were supported by the Institute for Basic Science in Korea (Grant No. IBS-R009-D1),
Samsung Science and Technology Foundation under Project Number SSTF-BA2002-06,
Basic Science Research Program through the National Research Foundation of Korea (NRF) (Grant No. 0426-20200003),
and the U.S. Army Research Office and and Asian Office of Aerospace Research \& Development (AOARD) under Grant Number W911NF-18-1-0137.
H.C.C was supported by the Institute for Basic Science in Korea (Grant No. IBS-R009-D1),
\\

\tbf{Author contributions} S.P. initially conceived the project. S.P. and Y.H. equally contributed to the theoretical analysis and wrote the manuscript with B.-J.Y.. H.C.C. did all of the ab initio calculations. B.-J.Y. supervised the project. All authors discussed and commented on the manuscript.
\\

\tbf{Competing financial interest statement} The authors have no competing financial interests to declare.

\clearpage
\section{Methods}

\subsection{Homotopy description of $\mf{q}$}
Here, we give a homotopy description of the topological charge $\mf{q}$.
As in the main text, we consider the $3\times 3$ real symmetric $H_{\bm{k}}$ at a fixed $k>0$, with the TATP situated at $k=0$.
Because there is a spectral gap between the $L$ mode and the $T$ modes for $k>0$ (note that there are two $T$ modes, $T_1$ and $T_2$), the Hamiltonian can be written as 
\ba
H_\bk=E^T_\bk \bpm
1 & 0 & 0 \\
0 & -1 & 0 \\
0 & 0 & -1 \epm
E_\bk, \quad
E_\bk=\bpm
\bep_{\bk,L} \\ \bep_{\bk,T_1} \\ \bep_{\bk,T_2} \epm,
\ea
after a spectral flattening in which the eigenvalues of the $L$ and the $T$ modes are sent to $1$ and $-1$, respectively.
Since $E_\bk \in O(3)$ and the Hamiltonian is invariant under $E_\bk \ra F_\bk E_\bk$ with $F_\bk \in O(1) \times O(2)$, the topological charge of the triple point at $\bk=0$ can be characterized by the second homotopy group~\cite{bzduvsek2017robust} of the classifying space $B=O(3)/[O(1) \times O(2)]$, which is $\pi_2(B)=2\mathbb{Z}$.
In the SI~\cite{supplement}, we use the exact sequence of homotopy groups for fibration to show explicitly that this topological charge is $2$ for isotropic phonons.
As further discussed in the SI~\cite{supplement}, this charge can be shown to be equivalent to the topological charge $\mf{q}=(\mf{n}_{sk},\mf{e})$ defined in the main text, where $\mf{e}=2\mf{n}_{sk}$, see also Refs.~\cite{bouhon2020non,bouhon2020geometric,unal2020topological}.

\subsection{Computation of Euler number using Wilson loop}
The absolute value of the Euler number $|\mf{e}|$ can be computed by the using the Wilson loop spectrum on a sphere surrounding the TATP.
To compute the Wilson loop spectrum, let us define the $2 \times 2$ overlap matrix  $[F_j]_{mn}=\bep_{\bk_j,m} \cdot \bep_{\bk_j,n}$, where $m,n \in \{ T_1, T_2\}$, and $\bk_j = k(\sin\theta \cos \phi_j, \sin \theta \sin \phi_j, \cos \theta)$ where $\phi = 2 \pi j/N$ for some integer $N$.
The Wilson loop operator at $\theta$ is defined as $W_\phi(\theta) = \lim_{N \ra \infty} F_{N-1}F_N...F_1F_0$.
Defining $w_{\phi}(\theta)$ to be the imaginary part the eigenvalues of $\ln W_\phi(\theta)$, we can compute $|\mf{e}|$ by counting the number of times $w_{\phi}(\theta)$ crosses $\pi$.

In the case of isotropic phonons, the transverse modes are tangent to the sphere on which $\mf{q}$ is defined, so that $\mf{e}=2$, and the Wilson loop spectrum shows the double winding structure. 
For isotropic phonons, there is an alternative explanation to this double winding structure in the Wilson loop spectrum.
Because $H_\bk$ is invariant under the $SO(2)$ rotation symmetry about the axis $\hat{\bk}$, we can split the eigenstates according to the eigenvalues of the helicity operator $\hat{\bk}\cdot \bb L$, where $\bb L=(L_x,L_y,L_z)$ is the spin 1 matrix representation of angular momentum. 
Because the helicity is $\pm 1$ for the transverse modes, we can define the Chern numbers for the transverse modes in the helicity sectors~\cite{supplement}, which are $\mp 2$.
Since the Wilson loop spectrum for a band with Chern number $n$ shows $n$ chiral windings, we see that there should be two branches with opposite winding in the Wilson loop spectrum, corresponding to the two helicity sectors with opposite Chern numbers.

\subsection{Vorticity and nodal lines}
We can define the vorticity of a Dirac point when the Hamiltonian has the $\mathcal{P}\mathcal{T}$ symmetry.
The effective Hamiltonian around a Dirac point can be written as a $2\times 2$ real symmetric matrix, $H_{D}=r(\bk) \cos \theta (\bk) \sg_x + r(\bk) \sin \theta (\bk) \sg_z$.
The vorticity is defined as the winding number of $(\cos \theta(\bm{k}), \sin \theta(\bm{k}))$ around the Dirac point. 
Although the vorticity can easily be defined locally around the Dirac point, its global definition is nontrivial.
A careful analysis~\cite{ahn2019failure,supplement} shows that a two-band insulator with Euler number $\mf{e}$ has even number of Dirac points such that the total sum of their vorticity $N_t$ satisfies the relation $-\tfrac{N_t}{2} = \mf{e}$.

This can be directly applied to the TATP: because the Euler number for the transverse acoustic phonons is $2$, there must be a minimum of four Dirac points on a sphere surrounding the ATP, such that the total sum of their vorticity is $-2\mf{e}$.
As we change the radius of this sphere, the trajectories of the Dirac points form nodal lines, so that there must be a minimum of four nodal lines emanating from the TATP.
Because two nodal lines emanating from the TATP can smoothly be connected, there must be a minimum of two nodal lines passing through the TATP (i.e. a nodal line emanating from the TATP is one half of a full nodal line passing through the TATP).

\subsection{Linked nodal structure protected by $\mf{q}$}
Let us explain why the symmetry protected TATP evolves into a nodal ring threaded by two nodal lines when the symmetry that protects the triple degeneracy is relaxed.
First, $\mf{n}_{sk}$ requires the gap between the $L$ mode and the $T$ modes to close inside the sphere on which $\mf{q}$ is defined.
However, because the triple point is no longer protected, and the generic nodal structure in a real symmetric Hamiltonian in 3D is the nodal line, the gap closing points between the $L$ mode and the $T$ modes evolve into a nodal ring, see \fig{Fig_3}d (see the SI~\cite{supplement} for the details of the Hamiltonian).
Second, $\mf{e}$ requires at least four Dirac points to form between the $T$ modes on the sphere on which $\mf{q}$ is defined. 
Equivalently, at least two nodal lines formed between the $T$ modes must pass through this sphere.
As can be seen in \fig{Fig_3}d, these two nodal lines formed between the $T$ modes (red lines) penetrate the nodal ring formed between the $L$ and the $T$ modes (black ring).
Such a structure is required because otherwise, it is possible for the nodal ring to be gapped out after deforming to a point, which is not compatible with the charge $\mf{n}_{sk}=1$ of the $L$ mode. 
We provide a simple geometric proof of this property in the SI~\cite{supplement}, and we note that a similar observation was also made in Ref.~[\onlinecite{tiwari2020non}] using quaternion charges.

\subsection{Details of ab initio calculations}
For the computation of the band structure and Wilson loop spectrum of the phonons in CsCl, we employed the Vienna ab initio simulation package (VASP)~\cite{kresse1996efficient} with the projector augmented-wave method (PAW)\cite{kresse1999ultrasoft}. 
The generalized gradient approximation (PBE-GGA) is employed for exchange-correlation potential~\cite{perdew1998perdew}. 
We used the default VASP potentials ($\textrm{Cs}_{\textrm{sv}}$ and Cl), and a 500 eV cutoff. 
To get the force constant, a $6 \times 6 \times 6$ supercell and a $6 \times 6 \times 6$ Monkhorst-pack k-point mesh were used. 
The phonon eigenvalues and the eigenstates were calculated using the PHONOPY package~\cite{togo2008first}. 
The dynamical matrix and the force constants were obtained from  the frozen phonon method, based on the Hellmann-Feynman theorem. 

Let us note that this calculation does not take into account the non-analytic correction terms to the dynamical matrix \cite{pick1970microscopic}.
Since the optical phonons in ionic insulators such as CsCl can be strongly renormalized by the non-analytic correction terms \cite{ahmad1972lattice,bingol2015electronic,he2017lattice}, the stability of the symmetry protected ATPs requires a more thorough analysis.

\let\addcontentsline\oldaddcontentsline
\clearpage
\onecolumngrid
\begin{center}
\bf \large Supplementary Information for ``\ourtitle''
\end{center}
\setcounter{section}{0}
\setcounter{equation}{0}
\setcounter{figure}{0}
\renewcommand{\thesection}{S\arabic{section}}
\renewcommand{\theequation}{S\arabic{equation}}
\renewcommand{\thefigure}{S\arabic{figure}}
\hfill \\
\twocolumngrid

\section{Proof that isotropic phonon is topological}

\begin{figure}[b]
\centering
\includegraphics[width=8.5cm]{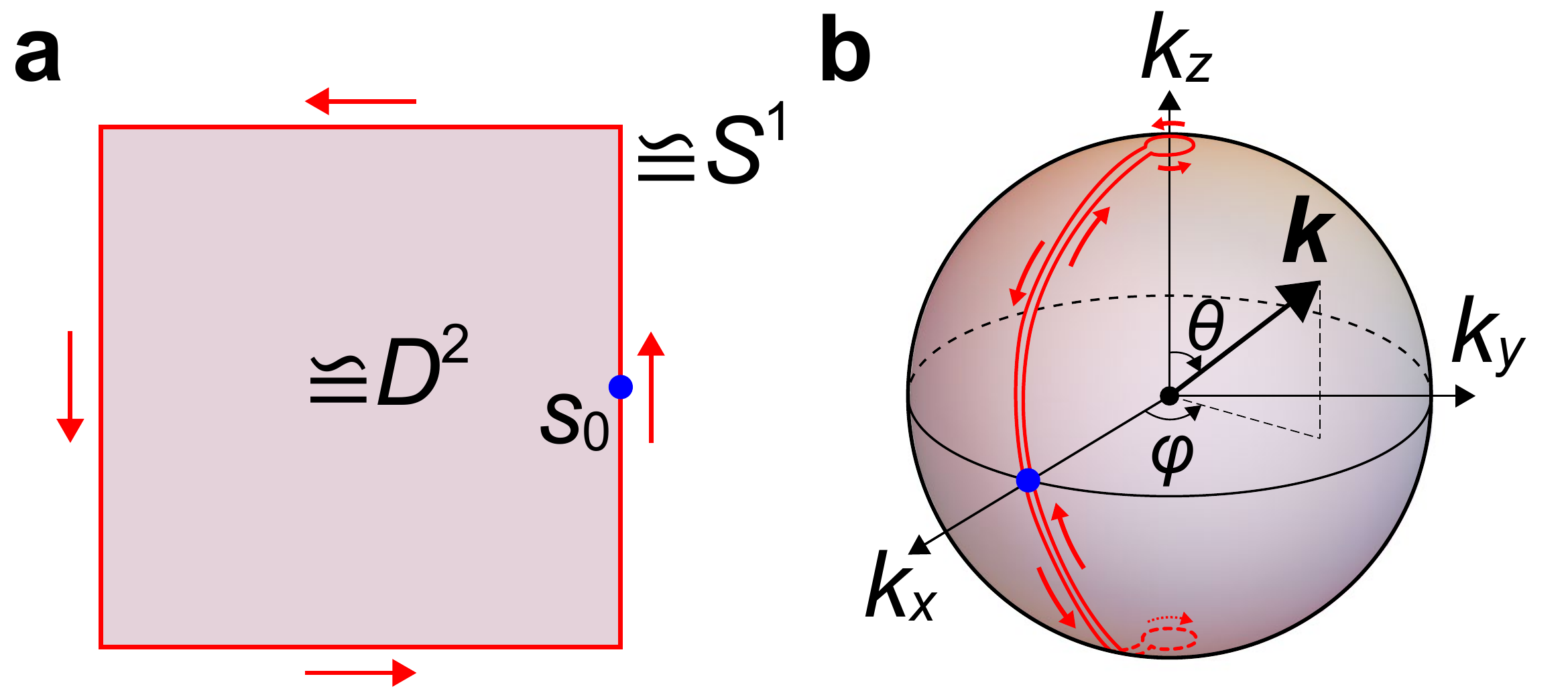}
\caption{\tbf{Evaluating the topological charge.} \tbf{a,} Illustration of $D^2$, its boundary $S^1$ (red line), and the base point $s_0$ (blue dot). 
The orientation of the boundary is indicated with the red arrows.
\tbf{b,} $D^2$ can be mapped to the sphere as shown.
The points to which $S^1$ is mapped is indicated by the red line with the orientation indicated by the red arrows.
}
\label{Fig_S1}
\end{figure}

In this section, we study the topological charge $\mf{q}$ in detail.
We also refer the readers to Refs.~\cite{bouhon2020non,bouhon2020geometric,unal2020topological}, where similar results were also obtained.
As in the main text, we use the dynamical matrix of isotropic phonon as the model Hamiltonian, whose matrix components are given by $[D_{\bk}]_{\alpha \beta} = v_T^2 k^2 \delta_{\alpha \beta} + (v_L^2 - v_T^2)k_{\alpha} k_{\beta}$.

In general, when the system has both the time reversal symmetry $\mc{T}$ ($\mc{T}^2=1$) and the inversion symmetry $\mc{P}$ ($\mc{P}^2=1$), or the combined symmetry $\mc{PT}$, the Hamiltonian can always be chosen to be real by choosing the gauge in which $\mc{PT}=\mc{K}$, where $\mc{K}$ is the complex conjugation operator.
Because there is also a gap between the longitudinal and the transverse modes away from $\bk=0$, the Hamiltonian can be written as $\bep^{T}_{\bk,L} \bep_{\bk, L} - \bep^{T}_{\bk,T_1} \bep_{\bk, T_1} - \bep^{T}_{\bk, T_2} \bep_{\bk, T_2}$ after sending the energy of the longitudinal mode ($L$) to $1$ and the transverse modes ($T_1$ and $T_2$) to $-1$. 
Equivalently,
\ba
H_{\bm{k}} =E^T_\bk \bpm
1 & 0 & 0 \\
0 & -1 & 0 \\
0 & 0 & -1 \epm
E_\bk, \quad
E_\bk = \bpm
\bep_{\bk,L} \\ \bep_{\bk,T_1} \\ \bep_{\bk, T_2} 
\epm.
\ea
Since $E_\bk \in O(3)$ and the Hamiltonian is invariant under $E_\bk \ra F_\bk E_\bk$ with $F_\bk \in O(1) \times O(2)$, the topological charge of the triple point at $\bk=0$ can be characterized by the second homotopy group of the classifying space $O(3)/[O(1) \times O(2)]$.

Because we choose a base point when computing the homotopy, we can assume that the classifying space $B$ is connected to the identity: $B=SO(3)/S[O(1) \times O(2)]$, where $S[O(1) \times O(2)]$ is the subgroup of $O(1) \times O(2)$ with unit determinant.
The space $SO(3)/S[O(1) \times O(2)]$ can be viewed as a fiber bundle with base space $B=SO(3)/S[O(1) \times O(2)]$, total space $E=SO(3)$, and fiber $F=S[O(1) \times O(2)]$. 
It is useful to note that $F$ consists of two types of matrices, once we choose a representation of $F$:
\ba
\bpm
1 & 0 & 0 \\
0 & \cos \theta & \sin \theta \\
0 & -\sin \theta & \cos \theta
\epm, \quad
\bpm
-1 & 0 & 0 \\
0 & - \cos \theta & -\sin \theta \\
0 & -\sin \theta & \cos \theta
\epm.
\ea
Therefore, $F$ has two connected components $F^+$ and $F^-$, characterized by the sign of the determinant of $O(1)$, or equivalently the sign of the determinant of $O(2)$, which is $\pm 1$, which corresponds to $F^\pm$.
We are interested in computing $\pi_n (B,b_0)$, where $b_0\in B$ is the base point.
To do this, we can examine the exact sequence for fibration~\cite{hatcher2002algebraic}:
\ba
\cdots \ra \pi_n (F,x_0) &\xrightarrow[]{} \pi_n(E,x_0) \xrightarrow[]{p_*} \pi_n (B,b_0) \nn \\
&\xrightarrow[]{} \pi_{n-1}(F,x_0) \ra \cdots \ra \pi_0(E,x_0),
\ea
where $p$ is a projection map $p: E \ra B$ and $x_0 \in F = p^{-1}(b_0)$.
Because we fix the base point when computing the homotopy group, we can take $F$ to be the component connected to the identity, i.e. $F^+$, which is topologically equivalent to the group $SO(2)$ (or equivalently, a circle).
We find it useful to note that the above sequence is actually a consequence of the following exact sequence for pairs because the projection map $p$ induces the isomorphism $p_*: \pi_n (E,F,x_0) \ra \pi_n(B,b_0)$ for $n\ge 1$:
\ba
\cdots \ra \pi_n (F,x_0) &\xrightarrow[]{i_*} \pi_n(E,x_0) \xrightarrow[]{j_*} \pi_n (E,F,x_0) \nn \\
&\xrightarrow[]{\der} \pi_{n-1}(F,x_0) \ra \cdots \ra \pi_0(E,x_0)
\ea
Here, the map $i_*$ and $j_*$ are induced by the inclusions $i:(F,x_0)\hookrightarrow (E,x_0)$ and $j:(X,x_0,x_0)\hookrightarrow (X,A,x_0)$.
The map $\der$ is induced by restricting the map $(D^n, S^{n-1}, s_0) \ra (X,A,x_0)$ to $S^{n-1}$, where $D^n$ is the $n$-dimensional disk, $S^{n-1}$ is the $(n-1)$-dimensional sphere, and $s_0$  is a point in $S^{n-1}$.

The part of the above sequence that we need is:
\ba
\cdots \pi_2(E,x_0) \xrightarrow[]{j_*} \pi_2(E,F,x_0) \xrightarrow[]{\der} \pi_1(F,x_0) \xrightarrow[]{i_*} \pi_1(E,x_0) \cdots
\ea
Using the homotopy data~\cite{ito1993encyclopedic} for $SO(n)$, this sequence becomes
\ba
\cdots 0 \xrightarrow[]{j_*} \pi_2(E,F,x_0) \xrightarrow[]{\der} \mathbb{Z} \xrightarrow[]{i_*} \mathbb{Z}_2 \cdots
\ea
Thus, $\pi_2(E,F,x_0) = 2\mathbb{Z}$, and $\der$ is an injective map of $\pi_2(E,F,x_0)$ into the kernel of $i_*$.
Thus, nontrivial elements of $\pi_2(E,F,x_0)$ can be characterized by maps $r:(D^2,S^1,s_0) \ra (E,F,x_0)$ such that $S^1$ is mapped to $F=SO(2)\cong S^1$ with two windings.

Using this, we can prove that the acoustic phonon modes are topologically nontrivial.
The unit normalized phonon polarization vectors are
\ba
\bep_{\bk,L} &= \frac{1}{k} (k_x,k_y,k_z) \\
\bep_{\bk,T_1} &=\frac{1}{\tilde{k}}(-k_y,k_x,0) \\
\bep_{\bk,T_2} &=\frac{1}{\tilde{k}k}(-k_x k_z,-k_yk_z,k_x^2+k_y^2).
\ea
The goal is to compute the topological charge $\pi_2(B,b_0) \cong\pi_2(E,F,x_0)$ defined on a sphere $S^2$ with origin at $\bk=0$ in the momentum space.
First, we note that 
\ba
E_\bk = \bpm
\bep_{\bk,L} \\
\bep_{\bk,T_1} \\
\bep_{\bk,T_2} 
\epm \in SO(3).
\ea
Since $D^2$ be identified with the sphere in the momentum space as illustrated in \fig{Fig_S1}a, $E_\bk$ can be viewed as a map from $D^2$ to $SO(3)/S[O(1) \times O(2)]$ which is well-defined everywhere on $D^2$.
(When we view $E_\bk$ as a map from the sphere to $E$, singularities arise at the N and S poles.)

Now, let us consider the following rotation matrix 
\ba
R_{\bk}(t) = \bpm
\cos t (\theta-\frac{\pi}{2}) & 0 & \sin t (\theta-\frac{\pi}{2}) \\
0 & 1 & 0 \\
-\sin t (\theta-\frac{\pi}{2}) & 0 & \cos t (\theta-\frac{\pi}{2})
\epm
\ea
where $\bk = k(\sin \theta \cos \phi, \sin \theta \sin \phi, \cos \theta)$.
Notice that $R_\bk(t)$ is nothing but the rotation about the $y$ axis by the angle $t(\theta-\frac{\pi}{2})$. 
When $t=0$, we see that it is the identity map, while for $t\neq 0$, it is a continuous function of $\theta$ on $S^2$.
Therefore, $E_\bk(t) = E_\bk R_\bk(t)$ defines a continuous deformation of an element in $\pi_{2}(B,b_0)$.
Here, we note that the energy gap between the longitudinal and the transverse modes is always preserved under this transformation.
Also, because the deformation is identity at $\theta=\pi/2$, the base point is preserved under this deformation if we choose $x_0$ to be the identity matrix in $SO(3)$, to which $s_0$ is mapped under $E_\bk$, see \fig{Fig_S1}. 
Now, $E_\bk(1) = E_\bk R_\bk(1)$ is an element of the component connected to the identity in $F$ as we trace along $S^1$ (the boundary of $D^2$), so that $E_\bk(1) \in \pi_2(E,F,x_0)$.
Therefore, the topological charge $\pi_{2}(B,x_0) \cong \pi_2(E,F,x_0)$ can be computed by counting the winding number in $F$ of the map $E_\bk(1)$ when $\bk$ is restricted to $S^1$ (the boundary of $D^2$): 
Near the N pole, we have
\ba
E_\bk(1) = \bpm
1 & 0 & 0 \\
0 & \cos \phi & \sin \phi \\
0 & -\sin \phi & \cos \phi
\epm,
\ea
and near the S pole, we have
\ba
E_\bk(1) = \bpm
1 & 0 & 0 \\
0 & \cos \phi & -\sin \phi \\
0 & \sin \phi & \cos \phi
\epm
\ea
while 
$E_\bk(1)$ is constant along the line connecting the north and the south poles.
Since the north pole is traversed counterclockwise while the south pole is traversed clockwise, we see that $S^1$ winds twice in $F\cong S^1$, i.e. the charge is $2 \in \pi_1(F,x_0)$.
As we explain below, this topological charge is nothing but the Euler number~\cite{ahn2018band,ahn2019failure} $\mf{e}$ of the transverse modes.
Let us also observe that the longitudinal mode has nonzero skyrmion number $\mf{n}_{sk}$.

To make the connection to the Euler number $\mf{e}$, let us begin by noting that the space $B=SO(3)/S[O(1) \times O(2)]$ can be thought of unoriented planes embedded in $\mathbb{R}^3$. 
Because all vector bundles over $S^2$ can be oriented~\cite{hatchervector}, we can choose a map from $S^2$ to $B$ to lie in $B^+=SO(3)/F^+=SO(3)/[SO(1) \times SO(2)]$, which is nothing but the space of oriented planes.
Alternatively, we can also choose the map to lie in $B^-=SO(3)/F^-$, which will be discuss later.
Since oriented planes are determined by an ordered pair of orthonormal vectors, the fiber bundle can be identified with the sphere bundle (fiber bundle with fiber $S^1$). 
It is well known that such bundles are characterized by the Euler number.
The Euler number can be computed by choosing a section, which can always be done over $S^2-\{x_1,...,x_k\}$ for finite non-negative integer $k$ (sphere minus finite number of points $x_i$), and counting the winding number around the points at which the section is not well-defined~\cite{bott2013differential}. 
This is essentially what we have done during the computation of $\pi_2 (B,x_0)$.
Another simple way to see that the transverse modes must be characterized by the Euler number is that the transverse modes form a basis for the tangent space to the sphere $S^2$.

Because we choose an ordered basis for the planes, normal vector to the oriented plane is fixed by the oriented plane through the cross product of the ordered basis (and vice versa).
The normal vector to the oriented plane is nothing but the line bundle, and because the line bundle on $S^2$ is always trivial~\cite{hatchervector}, we can choose a global section.
Because such a section is a map from $S^2$ (sphere in the momentum space) to $S^2$ (normalized longitudinal mode), its topological nature can be characterized by $\pi_2(S^2)=\mathbb{Z}$.
Since the skyrmion number of the longitudinal mode is $1$, we see that $\pi_2(B^+,x_0)$ is equivalent to twice the skyrmion number $\mf{n}_{sk}$ of the vector characterizing the oriented plane (that is, the longitudinal mode), and it is also equivalent to the Euler number $\mf{e}$ formed by the oriented planes: $\pi_2(B^+,x_0) = 2 \mf{n}_{sk} = \mf{e}$.

Although we have restricted the discussion above to $\pi_2(B^+,x_0)$, the Hamiltonian is in reality characterized by $\pi_2(B,x_0)$.
The difference here is that the orientation of the individual $O(1)$ and $O(2)$ sectors are not determined, and only the orientation of the $O(1) \times O(2)$ can be fixed.
Because we can always choose to orient the vector bundle, we conclude this section by carrying out a similar discussion for the topological charge $\pi_2(B^-,x_0)$.
This can easily be done by performing the transformation $E_\bk \ra E_\bk R_z(\pi)$, where
\ba
R_z(\pi) = \bpm
-1 & 0 & 0 \\
0 & -1 & 0 \\
0 & 0 & 1
\epm
\ea
is the rotation about the $z$ axis by $\pi$.
Then, the previous discussion on $\pi_2(B^+,x_0)$ applies without change except that $F^+$ is now replaced by $F^-$.
Since $R_z(\pi)$ reverses the orientation of the longitudinal mode and the transverse modes, the transformation reverses the signs of $\mf{e}$ and $\mf{n}_{sk}$.
Therefore, we see that by fixing an orientation for the vector bundles, we have $\pi_2(B,x_0)=\pm 2\mf{n}_{sk}=\pm \mf{e}$ (The sign is $+$ if we confine to $B^+$, while it is $-$ if we confine to $B^-$).
Therefore, the topological charge can be characterized by $\mf{q} = (\mf{n}_{sk},\mf{e})$, where $\mf{n}_{sk}$ is computed for the longitudinal mode, and $\mf{e}$ is computed for the transverse modes.

To conclude, because we can choose orientation for the fiber bundle in this case, the topological charge can be characterized by the Euler number, which is equivalent to twice the skyrmion number.
Also, although the sign of $\mf{n}_{sk}$ and $\mf{e}$ are not determinate in the sense that we can choose the fiber to lie in $B^+$ or $B^-$, there is no ambiguity in $\pi_{2}(B,x_0)$.

\section{Euler number as Chern numbers of helicity sectors}
As explained in the main text, we can explain the Euler number for the isotropic phonon Hamiltonian by computing the Chern numbers in the helicity sectors.
One way to do this is to introduce a Zeeman coupling along the $z$ direction, $V=L^z h$, where the $L^\rho$ are the usual angular momentum matrices:
\ba
L^x &= \bpm
0 & 0 & 0 \\
0 & 0 & -i \\
0 & i & 0
\epm, \\
L^y &= \bpm
0 & 0 & i \\
0 & 0 & 0 \\
-i & 0 & 0
\epm, \\
L^z &= \bpm
0 & -i & 0 \\
i & 0 & 0 \\
0 & 0 & 0
\epm.
\ea
Using the degenerate perturbation theory for the transverse modes, the zeroth order eigenstates are $\bep_{\bk,L}$, $\tfrac{1}{\sqrt{2}}(\bep_{\bk,T_1} + i \bep_{\bk,T_2})$, $\tfrac{1}{\sqrt{2}}(\bep_{\bk,T_1} - i \bep_{\bk,T_2})$.
These states are also eigenstates of the helicity operator $\hat{\bk} \cdot \bb L$ with eigenvalues given respectively by $0$, $1$, and $-1$.
Note that this lifts the degeneracy of the transverse modes, since the lowest corrections to the energy of the transverse modes with helicity $1$ and $-1$ are given by $E_T^{(0)}+h$ and $E_T^{(0)}-h$, where $E_T^{(0)}$ is the energy of the transverse modes without the perturbation.
It is straightforward to show that the Berry curvature for each of the helicity sectors with helicity eigenvalues $0$, $1$, and $-1$ are $0$, $-\tfrac{\hat{\bk}}{k^2}$, and $\tfrac{\hat{\bk}}{k^2}$, respectively.
Thus, the Chern numbers for each of the sectors are $0$, $-2$, and $2$, respectively.
Therefore, the Wilson loop spectrum for the $\pm 1$ helicity sectors should show a winding structure just like the state with $\mf{e}=2$.
Since this remains true in the limit $h \ra 0$, we see that the Wilson loop spectrum in the main text can be explained using Chern numbers of the helicity sectors.

\section{Elastic continuum Hamiltonian}
In this section, review the theory of elastic continuum \cite{nye1985physical,lifshitz1986theory} and give an example of the case with $\mf{q}=0$.

\subsection{Convention}
We first set down the conventions used in this work for the theory of elasticity.
First, the stress tensor $\sg_{ij}$ and the strain tensor $u_{ij}$ are related by elastic modulus tensor $\lambda_{ijkl}$ ($i,j,k,l=x,y,z$):
\ba
\sg_{ij}=\lambda_{ijkl} u_{kl}.
\label{eq.stress-strain}
\ea
Here, the strain and the stress tensors are symmetric,
\ba
u_{ij}=\frac{1}{2} \left( \frac{\der u_i}{\der x_j} + \frac{\der u_j}{\der x_i} \right), \quad \sg_{ij}=\sg_{ji}
\ea 
where $u_i$ is the displacement along the $i$th direction, and $x_i=(\bx)_i=(x,y,z)_i$.
The elastic modulus tensor satisfy
\ba
\lambda_{(ij)(kl)}=\lambda_{(kl)(ij)}, \quad \lambda_{(ji)(kl)}=\lambda_{(ij)(lk)}.
\ea
It follows that the elastic modulus tensor $\lambda_{ijkl}$ has a maximum of 21 independent components.
The potential energy is given by
\ba
U[u] = \frac{1}{2} \lambda_{ijkl} u_{ij} u_{kl}.
\ea

The dynamics of elastic system is described by 
\ba
\rho \frac{\der^2}{\der t^2} u_i = \frac{\der}{\der x_j} \sg_{ij}.
\ea
Its Fourier transformation is
\ba
\om^2_{\bm{k}} u_i(\bm{k})
= \left[ \rho^{-1} \lambda_{iljm} k_l k_m \right] u_j(\bm{k}),
\equiv D(\bk)_{ij} u_j(\bm{k}),
\label{eq.Christoffel_eq}
\ea
where $D(\bk)$ is called the dynamical matrix.
For notational simplicity, we absorb the mass density $\rho$ into the elastic modulus tensor, so that $D(\bk)_{ij} = \lambda_{iljm} k_l k_m$.
Let us note that in the main text, we used the notations $\mathcal{E}_{\bm{k}}=\om^2_{\bm{k}}$ and $H_{\bm{k}}=D(\bk)$.

We can simplify \eq{eq.stress-strain} with the help of the Voigt notation:
\begin{alignat}{4}
\sg_1 &= \sg_{xx}, \quad \sg_4 &&= \sg_{yz}, \quad \ep_1 &&= u_{xx} \quad \ep_4 &&= 2u_{yz}, \\
\sg_2 &= \sg_{yy}, \quad \sg_5 &&= \sg_{zx}, \quad \ep_2 &&= u_{yy} \quad \ep_5 &&= 2u_{zx}, \\
\sg_3 &= \sg_{zz}, \quad \sg_6 &&= \sg_{xy}, \quad \ep_3 &&= u_{zz} \quad \ep_6 &&= 2u_{xy}.
\end{alignat}
Now, \eq{eq.stress-strain} becomes $\sg_I = C_{IJ} \ep_J \quad (I,J=1,\dots,6)$.
In terms of $\lambda_{ijkl}$, the elastic tensor $C_{IJ}$ is expressed as
\ba
C_{IJ}=\bpm
\lambda_{xxxx} & \lambda_{xxyy} & \lambda_{xxzz} & \lambda_{xxyz} & \lambda_{xxxz} & \lambda_{xxxy} \\
& \lambda_{yyyy} & \lambda_{yyzz} & \lambda_{yyyz} & \lambda_{xzyy} & \lambda_{xyyy} \\
& & \lambda_{zzzz} & \lambda_{yzzz} & \lambda_{xzzz} & \lambda_{xyzz} \\
& & & \lambda_{yzyz} & \lambda_{xzyz} & \lambda_{xyyz} \\
& & & & \lambda_{xzxz} & \lambda_{xyxz} \\
& & & & & \lambda_{xyxy}
\epm_{IJ}.
\ea
Note that $C_{IJ}$ is symmetric, and we explicitly wrote down only its upper triangular part.

\begin{widetext}
\subsection{Symmetry properties}
For the classification of the elastic continuum Hamiltonian in a 3D crystal, it suffices to consider the 32 crystallographic point groups.
To find the constraints due to one of the point groups $G$, let the matrix representation of an element $\mc{G} \in G$ be $\tilde{G}$.
Then, $\lambda_{ijkl}$ satisfies
\ba
\lambda_{ijkl} = \tilde{G}_{im} \tilde{G}_{jn} \tilde{G}_{ko} \tilde{G}_{lp} \lambda_{mnop}.
\label{eq.lambda_sym}
\ea
By imposing the point group symmetries, it is known that there are 9 classes elastic tensors, see \Rf{nye1985physical}.
Here, we will focus on the trigonal and the cubic crystal systems.

\subsubsection{Trigonal}
The point groups $3=C_3$, $3m=C_{3v}$, $\overline{3}=S_6=C_{3i}$, $32=D_3$, and $\overline{3}m=D_{3d}$ belong to trigonal crystal system.
There are two classes of elastic tensor $C_{IJ}$ belonging to the trigonal crystal system, depending on the presence of either a twofold rotation symmetry or a mirror symmetry.

(i) Trigonal I: The point groups $3$ and $\overline{3}$ lack a twofold rotation symmetry or a mirror symmetry.
Then, $\lambda_{ijkl}$ has 7 independent elements, which can be organized using the Voigt notation as follows:
\ba
C_{IJ}=\bpm
C_{11} & C_{12} & C_{13} & C_{14} & C_{15} & \\
& C_{11} & C_{13} & -C_{14} & -C_{15} & \\
& & C_{33} & & & \\
& & & C_{44} & & -C_{15} \\
& & & & C_{44} & C_{14} \\
& & & & & \frac{1}{2}(C_{11}-C_{12})
\epm_{IJ}.
\ea

(ii)Trigonal II: The point groups $32$, $3m$, and $\overline{3}m$ have either a twofold rotation symmetry or a mirror symmetry, which kills $C_{15}$.
Then, $\lambda_{ijkl}$ has 6 independent elements, which can be organized using the Voigt notation as follows:
\ba
C_{IJ}=\bpm
C_{11} & C_{12} & C_{13} & C_{14} & & \\
& C_{11} & C_{13} & -C_{14} & & \\
& & C_{33} & & & \\
& & & C_{44} & & \\
& & & & C_{44} & C_{14} \\
& & & & & \frac{1}{2}(C_{11}-C_{12}) \label{eq.trig2}
\epm_{IJ}.
\ea

\subsubsection{Cubic}
The point groups $23=T$, $m3=T_h$, $\overline{4}3m=T_d$, $432=O$, and $m3m=O_h$ belong to cubic crystal system.
$\lambda_{ijkl}$ has 3 independent elements:
\ba
C_{IJ}=\bpm
C_{11} & C_{12} & C_{12} & & & \\
& C_{11} & C_{12} & & & \\
& & C_{11} & & & \\
& & & C_{44} & & \\
& & & & C_{44} & \\
& & & & & C_{44}
\epm_{IJ}.
\ea
For cubic crystal system, we explicitly write down the dynamical matrix $D(\bk)$,
\ba
D(\bk)_{ij} = \bpm
C_{11} k_x^2 +C_{44} (k_y^2+k_z^2) & (C_{12}+C_{44})k_x k_y & (C_{12}+C_{44}) k_z k_x \\
& C_{11} k_y^2 +C_{44} (k_z^2+k_x^2) & (C_{12}+C_{44})k_y k_z \\
& & C_{11} k_z^2 +C_{44} (k_x^2+k_y^2)
\epm. \label{eq.cubic_phonon}
\ea
\end{widetext}

\subsection{The case with $\mf{q}=(0,0)$}
In the main text, we have studied the acoustic phonons in cubic systems in detail.
We concluded there that there are two cases possible: either $\mf{q}=(1,2)$ or $\mf{q}$ is not defined.
However, when we lower the crystal symmetry, we find that it is possible to obtain $\mf{q}=(0,0)$.
Here, we study the acoustic phonon in tellurium crystal with space group P3${_1}$21 (space group number 152), with point group 32.
Thus, its elastic tensor takes the shape in \eq{eq.trig2}.
Using the data from Materials Project~\cite{de2015charting}, we can obtain the elastic continuum Hamiltonian.
The nodal structure of the energy spectrum was shown in Fig.~\magenta{5}a in the main text. 
Note that the nodal points lines occur only between the lower two energy bands.
Therefore, we can define the topological number $\mf{q}$, which can be computed (up to sign) from the Wilson loop spectrum, which was shown in Fig.~\magenta{5}b in the main text. 
Because the Wilson loop spectrum does not show any winding, $|\mf{e}|=0$.
As expected from the relation $\mf{e}=2\mf{n}_{sk}$, the longitudinal mode (highest energy mode) does not show any skyrmion texture, as was shown in Fig.~\magenta{5}c in the main text. 

\section{Continuum Hamiltonian}
In this section, we consider the general continuum Hamiltonian in the presence of the $O_h$ and the $T_h$ point groups.
This is done by expanding the Hamiltonian about the triple point using the Gell-Mann matrices, 
\ba
H_\bk=\sum_n \lambda_n f_n (\bk).\label{eq.effective_SM}
\ea
Because of the $\mc{PT}$ symmetry, only $\lambda_n$ with $n=1,3,4,6,8$ are relevant (real symmetric).
As before, we denote an element in a point group $G$ by $\mc{G}$, and we denote its matrix representation by $\tilde{G}$.
The constraint due to $\mc{G}$ is $\tilde{G} H_\bk \tilde{G}^{-1} = H_{\mc{G}\bk}$.
For notational convenience, we define $\Lambda=(\lambda_1,\lambda_3,\lambda_4,\lambda_6, \lambda_8)$.

\subsection{$O_h$ group: $T_{1u}$, $T_{1g}$ representations}
We find that the $T_{1g}$ representation gives the same constraints, so we explicitly work out only the $T_{1u}$ representation.
The transformation properties of $\Lambda$ are as follows:
\begin{align*}
\tilde{M}_x \Lambda \tilde{M}_x^{-1} &= (-\lambda_1,\lambda_3,-\lambda_4,\lambda_6,\lambda_8) \\
\tilde{M}_y \Lambda \tilde{M}_y^{-1} &= (-\lambda_1,\lambda_3,\lambda_4,-\lambda_6,\lambda_8) \\
\tilde{M}_z \Lambda \tilde{M}_z^{-1} &= (\lambda_1,\lambda_3,-\lambda_4,-\lambda_6, \lambda_8) \\
\tilde{C}_{4z} \Lambda \tilde{C}_{4z}^{-1} &= (-\lambda_1,-\lambda_3,\lambda_6,-\lambda_4, \lambda_8) \\
\tilde{C}_{4x} \Lambda \tilde{C}_{4x}^{-1} &= (\lambda_4,\tfrac{\lambda_3+\sqrt{3} \lambda_8}{2},-\lambda_1,-\lambda_6, \tfrac{3\lambda_3-\sqrt{3} \lambda_8}{2\sqrt{3}}).
\end{align*} 
The symmetry constrained $f_n$ are:
\ba
f_1 & =a k_x k_y \nn \\
f_3 & =b (k_x^2 - k_y^2) \nn \\
f_4 &= a k_x k_z \nn \\
f_6 &= a k_y k_z \nn \\
f_8 &= \tfrac{b}{\sqrt{3}}(k_x^2+k_y^2) - \tfrac{2b}{\sqrt{3}} k_z^2. \label{eq.Oh_T1u}
\ea
Let us note that this is equivalent to the cubic elastic continuum Hamiltonian in \eq{eq.cubic_phonon} once we subtract away the trace, with $a=C_{12}+C_{44}$ and $b=\tfrac{C_{11}}{2}-\tfrac{C_{44}}{2}$.
It is also useful to note that the isotropic elastic continuum Hamiltonian is obtained for $a=2b$.

\subsection{$O_h$ group: $T_{2u}$, $T_{2g}$ representations}
We first examine the $T_{2u}$ representation.
Here, we take the following matrix representation of the relevant group elements:
\ba
\tilde{M}_x = \bpm
-1 & 0 & 0 \\
0 & 1 & 0 \\
0 & 0 & 1
\epm \nn
\ea
\ba
\tilde{M}_y = \bpm
1 & 0 & 0 \\
0 & -1 & 0 \\
0 & 0 & 1
\epm \nn
\ea
\ba
\tilde{M}_z = \bpm
1 & 0 & 0 \\
0 & 1 & 0 \\
0 & 0 & -1
\epm \nn
\ea
\ba
\tilde{C}_{4z} = \bpm
0 & -1 & 0 \\
1 & 0 & 0 \\
0 & 0 & -1
\epm \nn
\ea
\ba
\tilde{C}_{4x} = \bpm
-1 & 0 & 0 \\
0 & 0 & 1 \\
0 & -1 & 0
\epm. \nn
\ea
The transformation properties of $\Lambda$ are as follows:
\begin{align*}
\tilde{M}_x \Lambda \tilde{M}_x^{-1} &= (-\lambda_1,\lambda_3,-\lambda_4,\lambda_6,\lambda_8) \\
\tilde{M}_y \Lambda \tilde{M}_y^{-1} &= (-\lambda_1,\lambda_3,\lambda_4,-\lambda_6,\lambda_8) \\
\tilde{M}_z \Lambda \tilde{M}_z^{-1} &= (\lambda_1,\lambda_3,-\lambda_4,-\lambda_6,\lambda_8) \\
\tilde{C}_{4z} \Lambda \tilde{C}_{4z}^{-1} &= (-\lambda_1,-\lambda_3,-\lambda_6,\lambda_4,\lambda_8) \\
\tilde{C}_{4x} \Lambda \tilde{C}_{4x}^{-1} &= (\lambda_4,\tfrac{\lambda_3+\sqrt{3} \lambda_8}{2},-\lambda_1,-\lambda_6, \tfrac{3\lambda_3-\sqrt{3} \lambda_8}{2\sqrt{3}}) .
\end{align*} 
The symmetry constrained $f_n$ are:
\ba
f_1 & =a k_x k_y \nn \\
f_3 & =b (k_x^2 - k_y^2) \nn \\
f_4 &= a k_x k_z \nn \\
f_6 &= -a k_y k_z \nn \\
f_8 &= \tfrac{b}{\sqrt{3}}(k_x^2+k_y^2) - \tfrac{2b}{\sqrt{3}} k_z^2 \label{eq.Oh_T2u}
\ea
Let us note that this Hamiltonian differs from that of $T_{1u}$ only by the transformations $a \ra -a$ and $k_x \ra -k_x$.

We next examine the $T_{2g}$ representation.
Here, we take the following matrix representation of the relevant group elements:
\ba
\tilde{M}_x = \bpm
-1 & 0 & 0 \\
0 & 1 & 0 \\
0 & 0 & -1
\epm \nn
\ea
\ba
\tilde{M}_y = \bpm
1 & 0 & 0 \\
0 & -1 & 0 \\
0 & 0 & -1
\epm \nn
\ea
\ba
\tilde{M}_z = \bpm
-1 & 0 & 0 \\
0 & -1 & 0 \\
0 & 0 & 1
\epm \nn
\ea
\ba
\tilde{C}_{4z} = \bpm
0 & -1 & 0 \\
1 & 0 & 0 \\
0 & 0 & -1
\epm \nn
\ea
\ba
\tilde{C}_{4x} = \bpm
0 & 0 & 1 \\
0 & -1 & 0 \\
-1 & 0 & 0
\epm. \nn
\ea
The action on the Gell-Mann matrices is
\begin{align*}
\tilde{M}_x \Lambda \tilde{M}_x^{-1} &= (-\lambda_1,\lambda_3,\lambda_4,-\lambda_6,\lambda_8) \\
\tilde{M}_y \Lambda \tilde{M}_y^{-1} &= (-\lambda_1,\lambda_3,-\lambda_4,\lambda_6,\lambda_8) \\
\tilde{M}_z \Lambda \tilde{M}_z^{-1} &= (\lambda_1,\lambda_3,-\lambda_4,-\lambda_6,\lambda_8) \\
\tilde{C}_{4z} \Lambda \tilde{C}_{4z}^{-1} &= (-\lambda_1,-\lambda_3,-\lambda_6,\lambda_4,\lambda_8) \\
\tilde{C}_{4x} \Lambda \tilde{C}_{4x}^{-1} &= (\lambda_6,\tfrac{\lambda_3-\sqrt{3} \lambda_8}{2},-\lambda_4,-\lambda_1,-\tfrac{\sqrt{3}\lambda_3+ \lambda_8}{2}). 
\end{align*} 
The symmetry constrained $f_n$ are:
\ba
f_1 & = a k_x k_y \nn \\
f_3 & =b (k_x^2 - k_y^2) \nn \\
f_4 &= a k_y k_z \nn \\
f_6 &= a k_x k_z \nn \\
f_8 &= -\tfrac{b}{\sqrt{3}}(k_x^2+k_y^2) + \tfrac{2b}{\sqrt{3}} k_z^2 .\label{eq.Oh_T2g}
\ea
Let us note that this Hamiltonian differs from that of the $T_{1u}$ representation only by the transformations $k_x \leftrightarrow k_y$ and $b \ra -b$.

\subsection{$T_h$ group: $T_{u}$, $T_{g}$ representations}
Next, let us examine the $T_h$ group.
Here, we note that although $T$, $T_h$, and $T_d$ groups all support representations with three-fold degeneracy, only the $T_h$ group has inversion symmetry.
Because the $T_u$ and the $T_g$ representations of the $T_h$ group yield the same continuum Hamiltonian, we explicitly work out only the $T_u$ representation.
We first consider the constraints from $T$ subgroup.
The matrix representations of the relevant symmetry elements are
\ba
\tilde{C}_{2x} = \bpm
1 & 0 & 0 \\
0 & -1 & 0 \\
0 & 0 & -1
\epm \nn
\ea
\ba
\tilde{C}_{2y} = \bpm
-1 & 0 & 0 \\
0 & 1 & 0 \\
0 & 0 & -1
\epm \nn
\ea
\ba
\tilde{C}_{2z} = \bpm
-1 & 0 & 0 \\
0 & -1 & 0 \\
0 & 0 & 1
\epm \nn
\ea
\ba
\tilde{C}_{3} = \bpm
0 & 0 & 1 \\
1 & 0 & 0 \\
0 & 1 & 0
\epm \nn
\ea
The transformation properties of $\Lambda$ are as follows:
\begin{align*}
\tilde{C}_{2x} \Lambda \tilde{C}_{2x}^{-1} &= (-\lambda_1,\lambda_3,-\lambda_4,\lambda_6,\lambda_8) \\
\tilde{C}_{2y} \Lambda \tilde{C}_{2y}^{-1} &= (-\lambda_1,\lambda_3,\lambda_4,-\lambda_6,\lambda_8) \\
\tilde{C}_{2z} \Lambda \tilde{C}_{2z}^{-1} &= (\lambda_1,\lambda_3,-\lambda_4,-\lambda_6,\lambda_8) \\
\tilde{C}_3 \Lambda \tilde{C}_3^{-1} &= (\lambda_6,\tfrac{-\lambda_3+\sqrt{3}\lambda_8}{2},\lambda_1,\lambda_4,-\tfrac{\sqrt{3}\lambda_3+\lambda_8}{2}).
\end{align*} 
The constraint is 
\ba
f_1 & =a k_x k_y \nn \\
f_3 & =b k_x^2 + c k_y^2 + d k_z^2 \nn \\
f_4 &= a k_x k_z \nn \\
f_6 &= a k_y k_z \nn \\
f_8 &= -\tfrac{2}{\sqrt{3}} [(\tfrac{b}{2}+c)k_x^2+(\tfrac{c}{2}+d)k_y^2+(\tfrac{d}{2}+b)k_z^2]. \label{eq.Th_f}
\ea
The $T_h$ group additionally has the $\mc{S}_6$ symmetry.
Its action is $(k_x,k_y,k_z)\rightarrow(-k_y,-k_z,-k_x)$ in the momentum space, so that its $T_u$ representation is 
\ba
\tilde{S}_6 = \bpm
0 & -1 & 0 \\
0 & 0 & -1 \\
-1 & 0 & 0
\epm. \nn
\ea
The transformation property of $\Lambda$ under $\mc{S}_6$ 
\ba
\tilde{S}_6 \Lambda \tilde{S}_6^{-1} &= (\lambda_4,-\tfrac{\lambda_3+\sqrt{3}\lambda_8}{2},\lambda_6,\lambda_1, \tfrac{\sqrt{3}\lambda_3-\lambda_8}{2}). \nn
\ea
Its constraint on $f_n$ is 
\ba
b+c+d=0. \label{eq.Th_condition}
\ea

Finally, we remark that although all of the cubic point group symmetries constrain the elastic continuum Hamiltonian in the same way, this is not true for the general continuum Hamiltonian.
Importantly, we see that the constraint due to $O_h$ and $T_h$ groups are not the same, whereas they give the same constraint on the elastic continuum Hamiltonian.


\section{Linking structure protected by $\mf{q}$}
\begin{figure}[t]
\centering
\includegraphics[width=8.5cm]{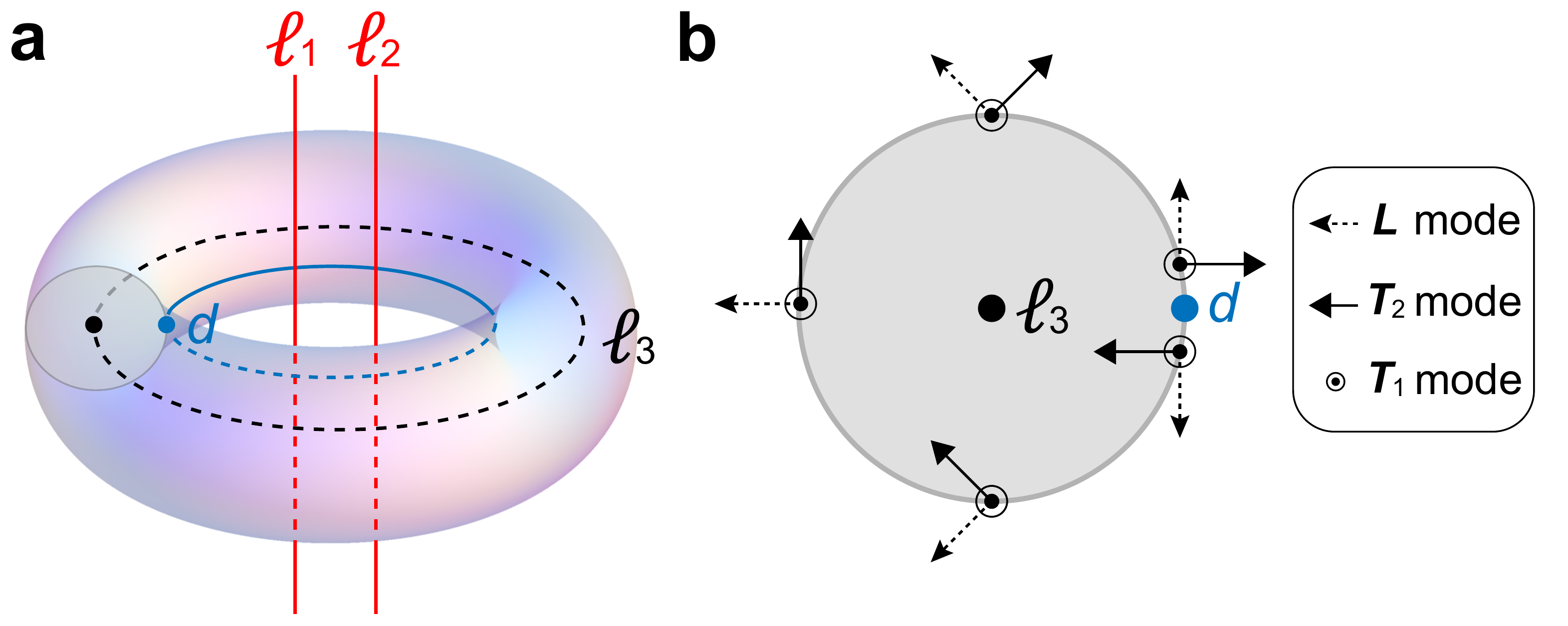}
\caption{\tbf{Linking structure.} \tbf{a,} $\ell_1$ and $\ell_2$ are the nodal lines formed between the $T_1$ and the $T_2$ modes. 
$\ell_3$ is the nodal ring formed between the $L$ and the $T_2$ modes is shown in black line.
$d$ is the ring of discontinuities in the $L$ mode on the torus.
\tbf{b,} A cross section of the torus in \tbf{a}.} 
\label{Fig_S2}
\end{figure}
In the main text, we claimed that when the triple point is perturbed such that the triple degeneracy is lifted while keeping the Hamiltonian components to be real, the resulting nodal structure is constrained by $\mf{q}$.
To demonstrate this, we plotted the nodal structure in Fig.~\magenta{3}b with the Hamiltonian of the form in \eq{eq.cubic_phonon} using the parameters $C_{11}=1.0$, $C_{12} = 0.6$, and $C_{44}=0.4$, so that $\mf{q}=(1,2)$. 
As a comparison with the case in which $\mf{q}$ is not defined, we plotted the nodal structure in Fig.~\magenta{3}c using the same form of the Hamiltonian with the parameters $C_{11}=1.0$, $C_{12} = 0.2$, and $C_{44}=2.0$. 
Both of these Hamiltonians were given a constant perturbation
\ba
\delta H = 0.02 \times \bpm
2 & -2 & 5 \\
-2 & -4 & 3\\
5 & 3 & -2
\epm
\ea
to plot the nodal structures in Fig.~\magenta{3}d,e.

The purpose of this section is to explain why TATP with $\mf{q}=(1,2)$ evolves into a nodal ring formed between the highest two bands ($L$ and $T_{2}$ modes) threaded by two nodal lines formed between the lowest two bands ($T_{2}$ and $T_{1}$ modes), as was seen in Fig.~\magenta{3}d.
For simplicity, we first assume that the TATP in the isotropic limit is perturbed by a term such as 
\ba
\delta H = 
\bpm
0 & 0 & 0 \\
0 & 0 & 0 \\
0 & 0 & \delta
\epm,
\ea
which preserves the cylindrical symmetry about the $z$ axis.
As summarized in \fig{Fig_S2}, let $\ell_1$ and $\ell_2$ (red lines) denote the two nodal lines formed between the $T_1$ and $T_2$ modes, and let $\ell_3$ (black ring) denote the nodal ring formed between the $L$ and the $T_2$ modes.
Note that we are assuming that $\mc{E}_{\bm{k},L} \geq \mc{E}_{\bm{k},T_2} \geq \mc{E}_{\bm{k},T_1}$. 
Although the lines $\ell_1$ and $\ell_2$ overlap in the momentum space due to the cylindrical symmetry, we draw them separately for clarity.

The goal is to show that $\ell_1$ and $\ell_2$ should penetrate $\ell_3$.
First, we note that $\ell_3$ is protected by the $\pi$-Berry phase.
Because of the $\pi$-Berry phase, $\bm{\ep}_{\bm{k},L}$ mode around the nodal ring $\ell_3$ shows winding structure as shown in \fig{Fig_S2}b.
Importantly, there is a discontinuity in $\bm{\ep}_{\bm{k},L}$ due to the $\pi$ Berry phase \cite{ahn2018band}, indicated by a blue dot and labeled as $d$ in \fig{Fig_S2}b.
Notice that this is compatible with the skyrmion texture of $\bm{\ep}_{\bm{k},L}$ on a sphere surrounding $\ell_3$.
In fact, because of the skyrmion texture, $\bm{\ep}_{\bm{k},L}$ on all 2D slices of the torus have similar wavefunction texture.
In particular, the wavefunction discontinuity indicated with a blue dot in \fig{Fig_S2}b forms a circle, as shown in \fig{Fig_S2}a as a blue line and labeled as $d$.

Now, because $\ell_3$ becomes a 2D Dirac point on the 2D slice (grey cut in \fig{Fig_S2}), the $\bm{\ep}_{\bm{k},T_2}$ has the texture schematically shown in \fig{Fig_S2}b.
Therefore, $\bm{\ep}_{\bm{k},T_1}$, being orthogonal to both $L_1$ and $T_2$ modes, is tangential to the nodal ring $\ell_3$.
It follows that just above the blue line, we have
\ba
\ep_{\bm{k},L} &= (0,0,1) \nonumber \\
\ep_{\bm{k},T_{2}} &= \tfrac{1}{\tilde{k}}(-k_x,-k_y,0) \nonumber \\
\ep_{\bm{k},T_{1}} &= \tfrac{1}{\tilde{k}}(-k_y,k_x,0).
\ea
Because the first two components of $T_{1}$ and $T_{2}$ modes have vorticity of $2$, by which we mean that they are eigenstates of the Hamiltonian of the form $H_{\bm{k}} \propto 2 k_x k_y \sigma_x + (k_x^2-k_y^2) \sigma_z$, there must be two Dirac points at $k_z=0$, which corresponds to the two nodal lines $\ell_1$ and $\ell_2$ threading the nodal ring $\ell_3$.

This geometric proof can also be generalized to the case in which the cylindrical symmetry is broken.
We only give a sketch of the proof since it does not give us further intuition.
We first consider a surface $D$ whose boundary is $\ell_3$.
Then, consider an arbitrarily small circular path $c$ surrounding $\ell_3$.
On this path, we choose gauge such that $\bm{\ep}_{\bm{k},L}$ has discontinuity at the point $d$ at which the circular path $c$ intersects the surface $D$.
Note that just above and below $d$, $\bm{\ep}_{\bm{k},L}$ point in the opposite directions due to the $\pi$ Berry phase provided by $\ell_3$.
Also, as we trace $c$, $\bm{\ep}_{\bm{k},L}$ and $\bm{\ep}_{\bm{k},T_2}$ must lie on the same plane: otherwise, $\bm{\ep}_{\bm{k},T_1}$, which is orthogonal to $\bm{\ep}_{\bm{k},L}$ and $\bm{\ep}_{\bm{k},T_2}$, will not converge to a single vector as we shrink the radius of $c$ to zero.
Now, let us consider $\bm{\ep}_{\bm{k},L}$ just above $d$. 
As we trace $\ell_3$, $\bm{\ep}_{\bm{k},L}$ traces a closed loop on a unit sphere. 
Now, consider a fixed orthonormal frame (a fixed set of orthonormal vectors) $\mc{F}_{\rm fix}$: ($\hat{\bm{x}}_{\rm fix}$, $\hat{\bm{y}}_{\rm fix}$, $\hat{\bm{z}}_{\rm fix}$).
We can define a local frame (orthonormal vectors as a function of $\bm{k}$ along the loop $\ell_3$) $\mc{F}_{\bm{k}, {\rm loc}}$: ($\hat{\bm{x}}_{\bm{k}, {\rm loc}}$, $\hat{\bm{y}}_{\bm{k}, {\rm loc}}$, $\hat{\bm{z}}_{\bm{k}, {\rm loc}}$) along $\ell_3$ by transforming $\mc{F}_{\rm fix}$ such that $\hat{\bm{z}}_{\bm{k}, {\rm loc}}$ aligns with $\bm{\ep}_{\bm{k},L}$ (e.g. rotate the fixed frame about the axis normal to $\bm{\ep}_{\bm{k},L}$ and $\hat{\bm{z}}_{\rm fix}$ to obtain $\mc{F}_{\bm{k}, {\rm loc}}$).
Then, the $\bm{\ep}_{\bm{k},T_1}$  and the $\bm{\ep}_{\bm{k},T_2}$ modes expressed in the local frame must have vorticity 2.
(Note that expressing $\bm{\ep}_{\bm{k},T_1}$ and $\bm{\ep}_{\bm{k},T_2}$ in this local frame is basically the same as deforming the $\bm{\ep}_{\bm{k},L}$ modes along $\ell_3$ just above $d$ to align with $\hat{\bm{z}}_{\rm fix}$ in the fixed frame. 
Since $\bm{\ep}_{\bm{k},L}$ mode has skyrmion texture, $\bm{\ep}_{\bm{k},T_1}$ and $\bm{\ep}_{\bm{k},T_2}$ must have vorticity $2$.)

We have thus shown that the skyrmion texture of the $L$ mode forces there to be Dirac points with total vorticity 2 in the ring $\ell_3$. 
There are many other ways to show why this should be true.
For example, we can similarly prove the linking structure by assuming that $\mf{e}=2$ for the $T_1$  and $T_2$ modes, and showing that the $L$ mode on the D-shaped closed path (e.g. begin from south pole in \fig{Fig_S1}, go straight to the north pole, and back to the south pole along the surface of the sphere) has $\pi$-Berry phase.
Alternatively, one can use the quaternion charge method as in \Rf{tiwari2020non}, or use the Dirac-string formulation in \Rf{ahn2019failure}.

\section{Nodal lines}

\subsection{Convention for counting nodal lines \label{appsub:nodallinenumber}}
Because of the Euler number $\mf{e}$ of the transverse modes defined on a sphere $S^2$ enclosing the triple point, there must be 2D Dirac points on $S^2$ with total vorticity of $-2\mf{e}$.
Because the low-energy Hamiltonian $\propto k^2$, these Dirac points form a nodal line in the 3D momentum space.
By the number of nodal lines \textit{emanating} from the triple point at $k=0$, we mean the number of Dirac points on the the sphere $S^2$.
Because two nodal lines emanating from the triple point can be naturally paired, when we refer to the number of nodal lines without any qualifications, we mean the number of paired nodal lines emanating from the triple point.
Note that when the dispersion of the Dirac points on the sphere is quadratic, the number of Dirac points is $2$.

Because of the relation $\mf{e}=-\tfrac{N_t}{2}$~\cite{ahn2019failure}, $\mf{e}$ places a constraint on the total vorticity $N_t$ of the Dirac points on $S^2$.
Here, the vorticity is defined by writing the effective Hamiltonian around the Dirac point as $H_{D}=r(\bk) \cos \theta (\bk) \sg_x + r(\bk) \sin \theta (\bk) \sg_z$, and counting the winding number of $(\cos \theta(\bm{k}), \sin \theta(\bm{k}))$.
Therefore, when $\mf{e} = 2$, there must be at least 4 Dirac points on the sphere, i.e. at least 4 nodal lines must emanate from the triple point.

\subsection{Nodal lines in continuum Hamiltonian of $O_h$ group and cubic elastic continuum}
\begin{figure}[t]
\centering
\includegraphics[width=8.5cm]{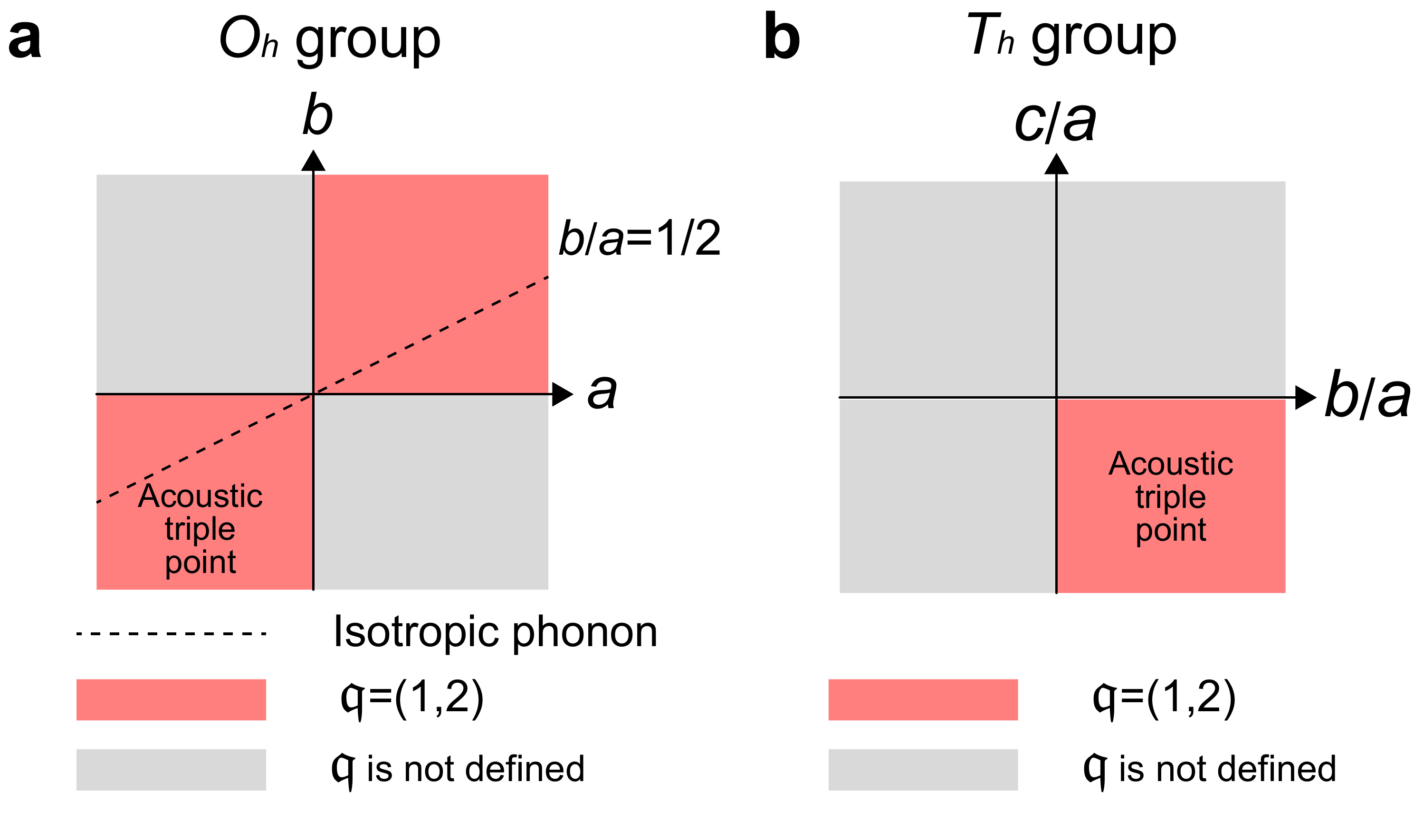}
\caption{\tbf{Topological phase diagram.} \tbf{a, b,} Phase diagram for point groups $O_h$ (\tbf{a}) and $T_h$ (\tbf{b}).}
\label{Fig_S3}
\end{figure}
Here, we study the nodal structure of the continuum Hamiltonian supporting three-dimensional representation for the $O_h$ group, or equivalently, the nodal structure in the elastic continuum Hamiltonian for the cubic crystal system.
Recall that the traceless Hamiltonian takes the form in \eq{eq.effective_SM} with $f_n$ given as in \eq{eq.Oh_T1u}.
Although it is not practical to analytically diagonalize the Hamiltonian at a generic $\bk$, we can gain an understanding of the nodal structure by diagonalize it along the high symmetry lines.
Now, we assume that $a\neq 0$, so that there is only one parameter $\tfrac{b}{a}$ that determines the form of the Hamiltonian.
Along the line with $k_y=k_z=0$, the eigenvalues are $\tfrac{4bk_x^2}{3a}$, $\tfrac{-2bk_x^2}{3a}$, $\tfrac{-2bk_x^2}{3}$, and along the line with $k_x=k_y=k_z$, the eigenvalues are $2 k_x^2$, $-k_x^2$, $-k_x^2$.
From this, we can conclude that for $\tfrac{b}{a}<0$, $\mf{q}$ cannot be defined because along the line with $k_x=k_y=k_z$, the lower two bands are degenerate, while for the line with $k_y=k_z=0$, the upper two bands are degenerate.
On the other hand, for $\tfrac{b}{a}>0$, the behavior along the high symmetry lines suggest that $\mf{q}$ can be defined, which we have confirmed through a numerical study in a reasonable parameter range.
Since we obtain the isotropic case for $b=\frac{a}{2}$, we conclude that $\mf{q}=(1,2)$ for $\tfrac{b}{a}\in\mathbb{R}^+$.
These results are summarized in \fig{Fig_S3}a.

\subsection{Nodal lines in $T_h$ group}
For the continuum Hamiltonian with the $T_u$ and $T_g$ representations of the $T_h$ group, the Hamiltonian is given by \eq{eq.effective_SM} with $f_n$ given by \eqs{eq.Th_f} and~\eqref{eq.Th_condition}.
As in the case with the $O_h$ group, we assume that $a \neq 0$, so that the Hamiltonian is determined by two parameters, $\tfrac{b}{a}$ and $\tfrac{c}{a}$ [Note that we can eliminate $d$ using \eq{eq.Th_condition}].
Since it is not practical to analytically diagonalize the Hamiltonian at a generic $\bk$, we first study the energy spectrum along the high symmetry lines.
Along $k_x=k_y=0$, the eigenvalues are $-\frac{2}{3a}(2b+c)k_z^2$, $\frac{2}{3a}(b+2c)k_z^2$, and $\frac{2}{3a} (b-c)k_z^2$. 
Along $k_x=k_y=k_z$, the eigenvalues are $2k_x^2$, $-k_x^2$, $-k_x^2$.
Thus, we expect the phase boundaries to be located at $\tfrac{b+c}{a}=0$, $\tfrac{b}{a}=0$, and $\tfrac{c}{a}=0$ (these are the parameters for which two of the energy bands along $k_x=k_y=k_z$ become equal).
Through a numerical study, we find that $\mf{q}=(1,2)$ when $\tfrac{b}{a}>0$ and $\tfrac{c}{a}<0$, as illustrated in \fig{Fig_S3}b.
Here, we note that this is consistent with the result for the $O_h$ group, since the the Hamiltonian reduces to that of the $O_h$ group (with $T_{1u}$ representation) when $c=-b$.

\section{Details of the 3D Lieb lattice  model}
In this section, we give the details of the 3D Lieb lattice model discussed in the main text.

\subsection{Tight-binding Hamiltonian}
The lattice structure is shown in \fig{Fig_4}a in the main text.
In each unit cell, there are 4 sublattice sites located at $\bx_1=(0,0,0)$, $\bx_2=(1/2,0,0)$, $\bx_3=(0,1/2,0)$, $\bx_4=(0,0,1/2)$.
The tight-binding Hamiltonian is given by
\ba
H_{\rm Lieb}(\bk)
=H_0(\bk) + H_1(\bk) + H_2 (\bk) + H_3 (\bk),
\ea
\ba
H_0(\bk)=\bpm
\ep_1 & 0 & 0 & 0 \\ 0 & \ep_2 & 0 & 0 \\ 0 & 0 & \ep_2 & 0 \\ 0 & 0 & 0 & \ep_2
\epm,
\ea
\ba
H_1(\bk)=2t_0 \bpm
0 & \cos\frac{k_x}{2} & \cos\frac{k_y}{2} & \cos\frac{k_z}{2} \\
\cos\frac{k_x}{2} & 0 & 0 & 0 \\
\cos\frac{k_y}{2} & 0 & 0 & 0 \\
\cos\frac{k_z}{2} & 0 & 0 & 0
\epm,
\ea
\ba
H_2(\bk)=4t_1 \bpm
0 & 0 & 0 & 0 \\
0 & 0 & \cos\frac{k_x}{2} \cos\frac{k_y}{2} & \cos\frac{k_z}{2} \cos\frac{k_x}{2} \\
0 & \cos\frac{k_x}{2}\cos\frac{k_y}{2} & 0 & \cos\frac{k_y}{2} \cos\frac{k_z}{2} \\
0 & \cos\frac{k_z}{2} \cos\frac{k_x}{2} & \cos\frac{k_y}{2} \cos\frac{k_z}{2} & 0
\epm,
\ea
\ba
H_3(\bk)=2 \left( \cos k_x + \cos k_y + \cos k_z \right) \bpm
t_2 & 0 & 0 & 0 \\ 0 & t_3 & 0 & 0 \\ 0 & 0 & t_3 & 0 \\ 0 & 0 & 0 & t_3 \epm.
\label{eq.H_Lieb}
\ea
Here, $\ep_1$ is the onsite potential for the s orbital located at $\bx_1$, and $\ep_2$ is the onsite potential for the s orbitals located at $\bx_2$, $\bx_3$, and $\bx_4$.
$t_0$, $t_1$, and $t_2$ are the hopping amplitudes for the nearest, the second-nearest, and the third-nearest neighbors.
We note that for the band structure in the main text, we have chosen the parameters $\ep_1=-2.0$, $\ep_2=1.2$, $t_0=1.0$, $t_1=0.3$, $t_2=0.2$, and $t_3=0.1$.

\subsection{Effective Hamiltonian of the triple point}
The tight-binding Hamiltonian $H_{\rm Lieb}(\bk)$ has the $O_h$ point group symmetry and the time-reversal symmetry $\mc{T}$.
As was noted in the main text, the band structure exhibits a triple point at $R=(\pi,\pi,\pi)$.
This triple point is protected by the $C_{2x}$, $C_{2y}$, and $C_{3[111]}$ symmetries~\cite{bradlyn2016beyond}.

By making an analogy between the triple point at $R$ and the acoustic phonon, we can map the highest energy band to the longitudinal phonon mode and the lower two energy bands to the transverse phonon modes by explicitly computing the effective Hamiltonian $H_{\rm eff}(\bk)$ near the triple point using the L{\"o}wdin perturbation theory~\cite{lowdin1951note}:
\begin{widetext}
\ba
H_{\rm eff}(\bk)_{nm}
= h(\bk_0)_{nm}
+ \sum_{i=1}^3 h_i(\bk_0)_{nm} k_i
+ \sum_{i,j=1}^3 \left[ \frac{1}{2} h_{ij}(\bk_0)_{nm} + \sum_{\overline{m} \ne n,m} \frac{h_i(\bk_0)_{n\overline{m}} h_j(\bk_0)_{\overline{m}m}}{E_n(\bk_0)-E_{\overline{m}}(\bk_0)} \right] k_i k_j + O(k^3),
\ea
\ba
h(\bk_0)_{ab}=\bra{a, \bk_0} H_{\rm Lieb}(\bk_0) \ket{b, \bk_0}, \quad h_i(\bk_0)_{ab}=\bra{a, \bk_0} (\der_i H_{\rm Lieb})(\bk_0) \ket{b, \bk_0}, \quad h_{ij}(\bk_0)_{ab}=\bra{a, \bk_0} (\der_i \der_j H_{\rm Lieb})(\bk_0) \ket{b, \bk_0},
\ea
where $\bk_0=(\pi,\pi,\pi)$, and the indices $n,m$ run over the bands that form the triple point, while $\overline{m}$ runs over the other bands, the lowest band in this case.
Also, the band indices $a,b$ run over any band.

Straightforward computation yields
\ba
H_{\rm eff}(\bk)
&=\bpm
t_3(k^2-6) + \frac{t_0^2}{6t_2-\ep_1} k_x^2 + \ep_2 & (t_1+\frac{t_0^2}{6t_2-\ep_1}) k_x k_y & (t_1+\frac{t_0^2}{6t_2-\ep_1}) k_z k_x \\
(t_1+\frac{t_0^2}{6t_2-\ep_1}) k_x k_y & t_3(k^2-6) + \frac{t_0^2}{6t_2-\ep_1} k_y^2 + \ep_2 & (t_1+\frac{t_0^2}{6t_2-\ep_1}) k_y k_z \\
(t_1+\frac{t_0^2}{6t_2-\ep_1}) k_z k_x & (t_1+\frac{t_0^2}{6t_2-\ep_1}) k_y k_z & t_3(k^2-6) + \frac{t_0^2}{6t_2-\ep_1} k_z^2 + \ep_2
\epm \\
&= \left(t_3(k^2-6)+\ep_2\right) \mathds{1}_3 - t_1 {\rm Diag}(k_x^2,k_y^2,k_z^2) + \left(t_1+\frac{t_0^2}{6t_2-\ep_1}\right) (k_x, k_y, k_z)^T (k_x, k_y, k_z).
\label{eq.effH}
\ea
We can therefore decompose $H_{\rm eff}(\bk)$ as
\bg
H_{\rm eff}(\bk)
= (\ep_2-6t_3) \mathds{1}_3 + D_{\rm eff}(\bk), \\
D_{\rm eff}(\bk) = \bpm
C_{11} k_x^2 +C_{44} (k_y^2+k_z^2) & (C_{12}+C_{44})k_x k_y & (C_{12}+C_{44}) k_z k_x \\
(C_{12}+C_{44})k_x k_y & C_{11} k_y^2 +C_{44} (k_z^2+k_x^2) & (C_{12}+C_{44})k_y k_z \\
(C_{12}+C_{44}) k_z k_x & (C_{12}+C_{44})k_y k_z & C_{11} k_z^2 +C_{44} (k_x^2+k_y^2) \epm,
\eg
where
\ba
C_{11} = \frac{t_0^2}{6t_2-\ep_1}+t_3, \quad C_{12} = t_1+\frac{t_0^2}{6t_2-\ep_1}-t_3, \quad C_{44} = t_3.
\ea
It is crucial to note that $D_{\rm eff}(\bk)$ is identical to the dynamical matrix in cubic symmetric system.
In this way, we can directly compare the triple point in the electronic system and that in the elastic material.
\end{widetext}

\section{The effect of boundary condition on surface localized states \label{sec.bc}}
In this section, we first study the implications of Dirichlet boundary condition, which is sometimes employed when studying the surface states of topological insulators.
As a comparison, we also review the free boundary condition which is known to yield surface acoustic waves.

\subsection{Dirichlet boundary condition}
Since there is no direct analog of the `free boundary condition' of phonons in electronic systems, we first study the Dirichlet boundary condition.
In the bulk, the equation of motion is
\ba
i\frac{\der}{\der t} \psi = (-a \nabla^2 - b \nabla^T \nabla) \psi .\label{eq.electron_eom}
\ea
We introduce a boundary at $z=0$ and study waves propagating along the $k_x$ direction  ($k_y=0$).
To solve this problem, let us first impose some symmetry constraints.
The mirror symmetry
\ba
M_y = \bpm
1 & 0 & 0 \\
0 & -1 & 0 \\
0 & 0 & 1
\epm
\ea
gives $\psi_y = 0$.
Further, we impose 
\ba
C_{2z}T = \bpm
1 & 0 & 0 \\
0 & 1 & 0 \\
0 & 0 & -1
\epm \mc{K},
\ea
so that $\psi_x$ is purely real and $\psi_z$ is purely imaginary.
Then, the wavefunction naturally satisfy the Hermiticity condition, which is often imposed when solving for the Fermi arc of Weyl semimetal:
\ba
0 =& \big[ a \left( \langle \nabla_z \psi_2, \psi_1 \rangle - \langle \psi_2, \nabla_z \psi_1 \rangle \right) \nn \\
&+ b \left( \nabla_z \psi_{2z}^* \psi_{1z} - \psi_{2z}^* \nabla_z \psi_{1z} \right) \nn \\
&- i b k_x \left( \psi_{2z}^* \psi_{1x} + \psi_{2x}^* \psi_{1z} \right) \big]_{z=0}.
\ea
It is convenient separate the wavefunction into the transverse and the longitudinal modes. 
Note that this is possible since we can expand any wave function by linear combination of the transverse and the longitudinal modes.
For the transverse part we write
\ba
\psi_T = \bpm \psi_{Tx} \\ \psi_{Ty} \\ \psi_{Tz} \epm
e^{\kappa_T z} e^{i k_x x - i \omega t},
\ea
where $\kappa_T$ is the inverse decay length, and $\psi_{T(x,y,z)}$ are constants.
The transversality condition $\nabla \cdot \psi_T=0$ gives
\ba
ik_x \psi_{Tx} + \kappa_T \psi_{Tz}=0 .
\ea
Also, the equation of motion in the bulk gives
\ba
\kappa_T = \sqrt{k_x^2 - \omega/a} .
\ea
Thus,
\ba
\psi_T = c_T \bpm \kappa_T \\ 0 \\ -ik_x \epm
e^{i k_x x -i\omega t}e^{\kappa_T z} . \label{eq.transverse_psi}
\ea

Similarly, the longitudinal part is written as
\ba
\psi_L = \bpm \psi_{Lx} \\ \psi_{Ly} \\ \psi_{Lz} \epm
e^{\kappa_L z} e^{i k_x x - i \omega t} .
\ea
The longitudinal condition $\nabla \times \psi_L=0$ gives
\ba
\kappa \psi_{Lx} - i k_x \psi_{Lz}=0.
\ea
Also, the equation of motion in the bulk gives
\ba
\kappa_L = \sqrt{k_x^2 - \omega/(a+b)}.
\ea
Thus,
\ba
\psi_L = c_L \bpm k_x \\ 0 \\ -i \kappa_L \epm
e^{i k_x x -i\omega t}e^{\kappa_L z} . \label{eq.longitudinal_psi}
\ea

The Dirichlet condition is that $\psi=0$ at the boundary.
This requires the following two equations to be satisfied:
\ba
c_T \sqrt{k_x^2-\omega/a} + c_L k_x &= 0 \\
c_T k_x + c_L \sqrt{k_x^2 - \omega/(a+b)} &= 0. \label{eq.dirichlet}
\ea
Thus, we must have
\ba
-\frac{k_x^2}{\sqrt{k_x^2 - \omega/a}} + \sqrt{k_x^2 - \omega/(a+b)}=0.
\ea
This yields two solutions
\ba
\omega &= 0 \nn \\ 
\omega &= k_x^2 (2a+b).
\ea
When the first condition is satisfied, \eq{eq.dirichlet} demands $k_x=0$, so that the solution is not really independent of $\omega = k_x^2(2a+b)$.

\begin{figure}[t]
\centering
\includegraphics[width=8.5cm]{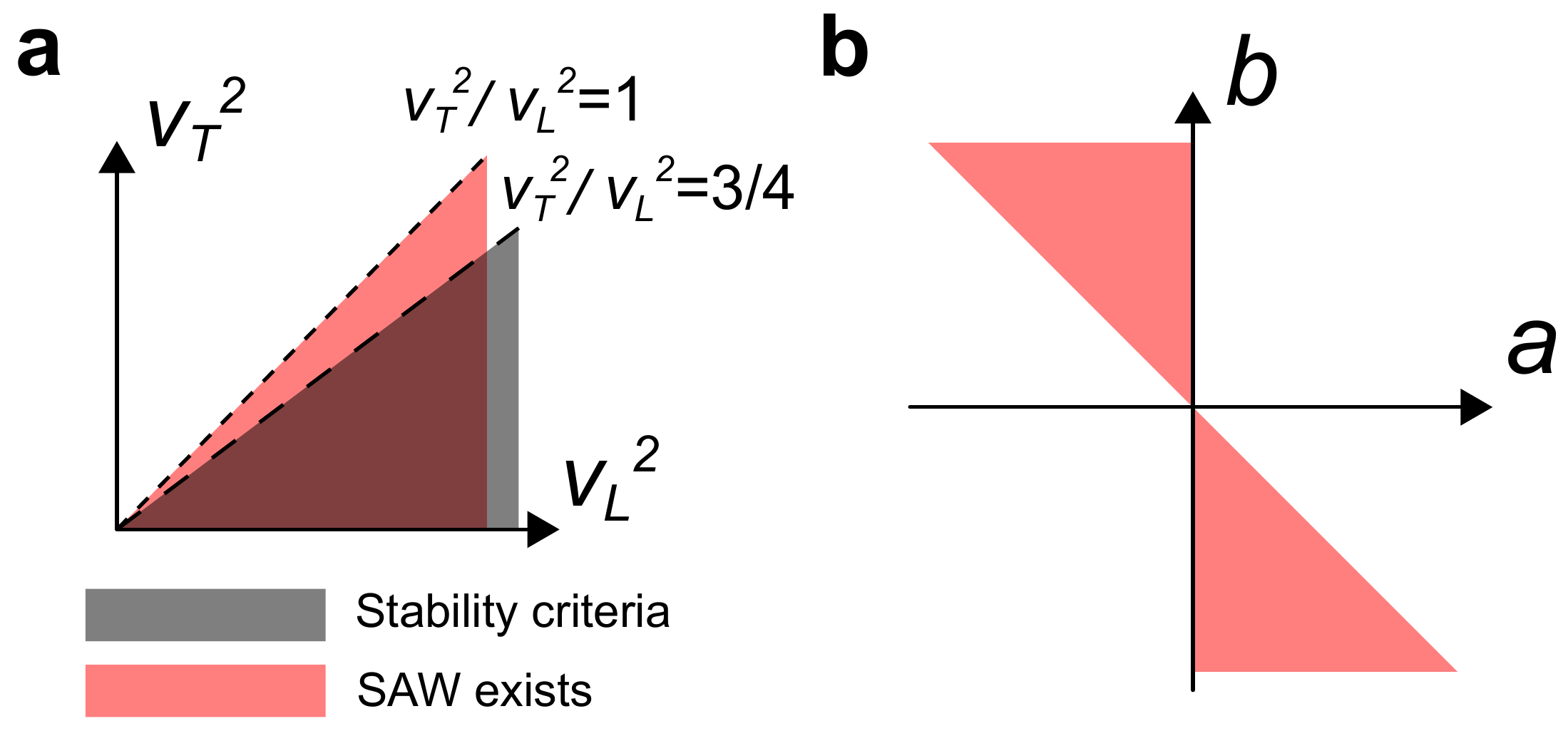}
\caption{\tbf{Conditions for the presence of surface localized states.}
\tbf{a,} Diagram showing when surface localized states can appear for free boundary condition. 
Note that when phonon satisfy the stability criteria, surface acoustic wave always appear.
\tbf{b,} Diagram showing when surface localized states can appear for Dirichlet boundary condition.
}
\label{Fig_S4}
\end{figure}

Let us note that the following conditions must be satisfied to obtain surface-localized states:
\ba
-\frac{b}{a} & > 1 \nn \\
\frac{b}{a+b} & > 1. \label{eq.dirichlet_local}
\ea
The parameter region satisfying these conditions is shown in \fig{Fig_S4}b.

Although the Dirichlet boundary condition is sometimes used to show that surface localized states exist in topological insulators, we note that naively solving for the wavefunctions in a finite size system does not yield surface modes consistent with the Dirichlet boundary condition.
To illustrate this, let us study the Lieb lattice with the parameters chosen such that the triple point at $R$ falls in the region in which the Dirichlet boundary condition yields surface localized modes, see \eq{eq.dirichlet_local}.

\begin{figure}[t]
\centering
\includegraphics[width=8.5cm]{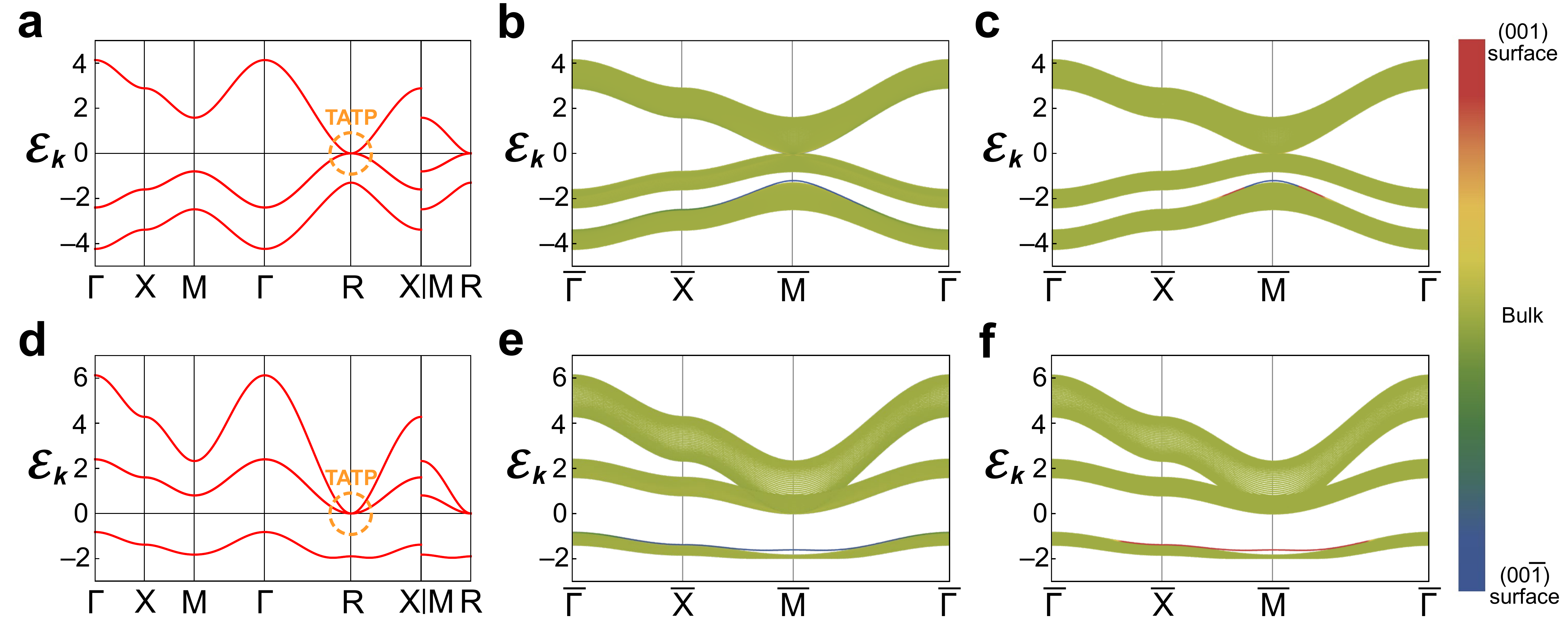}
\caption{
\tbf{Absence of surface states in 3D Lieb model.}
\tbf{a,} The band structure of the 3D Lieb model in the isotropic limit with the parameters $\ep_1=0.5$, $\ep_2=-1.2$, $t_0=1.0$, $t_1=0.0$, $t_2=0.3$, and $t_3=-0.2$. Note that two bands corresponding to transverse phonon modes are degenerate. The parameters are chosen such that if we impose the Dirichlet boundary condition, surface localized modes appear.
\tbf{b,} Surface spectrum for 001 termination. Color indicates the center of wave function. Note that the finite-size slab is constructed by stacking 20 layers along the $z$ direction.
\tbf{c,} Same as \tbf{b} but we remove the dangling atoms so that the finite-size slab is mirror symmetric in the $z$-direction.
\tbf{d,} The band structure with the $\ep_1=0.5$, $t_0=1.0$, $t_1=0.0$, $t_2=0.4$, $t_3=0.2$.
The parameters are chosen that $a=0.2$, $b=0.5263$ in \eq{eq.phonon_eom} to mimic the isotropic phonon. Notice that there are no surface localized modes analogous to the surface acoustic waves in \tbf{e} and \tbf{f}, which are the surface spectrums for asymmetric and symmetric termination, respectively, in the 001 direction.
}
\label{Fig_S5}
\end{figure}

We first notice from \eq{eq.effH} that the isotropic limit can be achieved by setting $t_1=0$ and $\ep_2=6t_3$.
In order to test the existence of surface states, we choose the parameter values,
$\ep_1=0.5, \, \ep_2=-1.2, \, t_0=1.0, \, t_1=0.0, \, t_2=0.3, \, t_3=-0.2$.
The band structure along the high symmetry lines is shown in \fig{Fig_S5}a.
We note that the second and third lowest bands are degenerate throughout the BZ.
The continuum Hamiltonian near $R$ is given by
\ba
H(\bk)=-0.2 k^2 \mathds{1}_3 + 0.77 (k_x, k_y, k_z)^T (k_x, k_y, k_z),
\ea
which corresponds to \eq{eq.electron_eom} with $a=-0.2$ and $b=0.77$. From \eq{eq.dirichlet_local}, we expect surface states for Dirichlet boundary condition.
However, we do not observe the desired surface states in the surface spectrum with both asymmetric and symmetric terminations (\fig{Fig_S5}b,c).
In fact, it is not clear how Dirichlet boundary condition can be achieved for discrete tight-binding model.

We note in passing that for phonon, the equation of motion is
\ba
-\frac{\der^2 \psi}{\der t^2} = (-a \nabla^2 - b \nabla^T \nabla) \psi. \label{eq.phonon_eom}
\ea
Thus, the only difference is that $\omega \ra \omega^2$, and $a = v_T^2$, $b = v_L^2-v_T^2$. 
We see that there are no localized states for the Dirichlet boundary condition for phonons (note that we always have $v_L>v_T$ for isotropic phonon).

\subsection{Free boundary condition \label{ssec.fbc}}
For completeness, we show that surface acoustic waves appear in isotropic medium by following \Rf{lifshitz1986theory}.
It is well known that isotropic phonons support surface acoustic waves. 
Denoting the stress tensor by $\sg_{ij}$, the free boundary condition is $\sum_k \sg_{ij} n_j=0$, where $\bb n= (0,0,1)$, since we are assuming a boundary at $z=0$. 
This entails the following: $\sg_{xz}=\sg_{yz}=\sg_{zz}=0$.
For isotropic medium, we have $\sg_{ij} = K u_{kk}\delta_{ij}+2\mu(u_{ij}-\frac{1}{3} \delta_{ij}u_{kk})$, where $u_{ij}=\frac{1}{2} (\frac{\der u_i}{\der x_j} + \frac{\der u_j}{\der x_i})$ is the strain tensor.
Here, the $K$ and $\mu$ are related to the longitudinal and transverse velocity by the relations $K=\lambda+\frac{2}{3}\mu$, $\mu=v_T^2$, and $\lambda = v_L^2-2v_T^2$.
Thus, $\sg_{yz}=0$ implies $u_{yz}=0$. 
Since we assume the that the wave does not depend on the value of $y$, this implies $\frac{\der u_y}{\der z}=0$. 
Using the surface wave ansatz, we obtain $u_y=0$.

Next, we note that the transverse and longitudinal modes essentially takes the form in \eqs{eq.transverse_psi} and \eqref{eq.longitudinal_psi}, respectively. 
With this in mind, we now examine the other two constraints.
The condition $\sg_{xz}=0$ implies $u_{xz}=0$, which translates to $c_T (k_x^2+\kappa_T^2) + 2c_L k_x \kappa_L=0$.
The condition $\sg_{zz}=0$ gives $2c_T v_T^2 \kappa_T k_x + c_L(2 v_T^2 k_x^2 +v_L^2(\kappa_L^2-k_x^2)) =0$.
These two combines to $4k_x^2 \kappa_L \kappa_T = (k_x^2 + \kappa_T^2)^2$. 
Thus, $\omega$ and $k_x$ satisfy $\frac{\omega^8}{v_T^8}-8k_x^2 \frac{\omega^6}{v_T^6}+8 k_x^4 \frac{\omega^4}{v_T^4}(3-2\frac{v_T^2}{v_L^2})+16k_x^6 \frac{\omega^2}{v_T^2}(\frac{v_T^2}{v_L^2}-1) =0$.
We then use the ansatz $\omega = v_T k_x \xi$ to obtain $\xi^6-8\xi^4+8\xi^2(3-2\frac{v_T^2}{v_L^2})+16(\frac{v_T^2}{v_L^2}-1) =0$.

Let us note that the localization requires $1-\xi^2>0$ and $1-\frac{v_T^2}{v_L^2}\xi^2>0$.
Also, the SAW at $(k_x',k_y')$ is $R(\phi) u_{\rm SAW} (\bk')$, where $\bk' = R(\phi)\bk$, where $R$ is rotation by $\phi$ about the $z$ axis.
For isotropic case, the longitudinal velocity always exceeds the transverse velocity due to the Born stability criterion that $v_T^2/v_L^2 < \tfrac{3}{4}$.
Since the SAW exists for $v_T^2/v_L^2 < 1$, we see that isotropic medium always have SAW for free boundary, as summarized in \fig{Fig_S4}a.
However, this may not be true when we move away from isotropic case, and it would be interesting to investigate the properties of surface acoustic waves in relation to the topology.

Finally, let us note because the equation of motion for phonon is $\frac{\der^2 \bu_\bk}{\der t^2} = -H(\bk) \bu_\bk$ and the equation of motion electron is $i\frac{\der \psi_\bk}{\der t} = H(\bk)\psi_\bk$, the wavefunction are eigenvectors $H(\bk)$.
However, the free boundary condition for phonon does not have a clear interpretation for electrons, and we are not guaranteed to obtain surface modes.
We demonstrate this using the Lieb lattice example.
As before, we choose the parameters such that the continuum Hamiltonian near $R$ can be mapped to isotropic phonon.
Here, we additionally demand $v_L>v_T>0$ under this identification.
The parameters $\ep_1=0.5$, $t_0=1.0$, $t_1=0.0$, $t_2=0.4$, $t_3=0.2$ yield $a=0.2$, $b=0.5263$ in \eq{eq.phonon_eom}.
The energy bands along the high symmetry lines are shown in \fig{Fig_S5}d.
Although the isotropic phonon should show surface localized states, we do not find any for the electronic case, as can be seen in \fig{Fig_S5}e,f for both asymmetric and symmetric terminations.

\begin{figure*}[t]
\centering
\includegraphics[width=17cm]{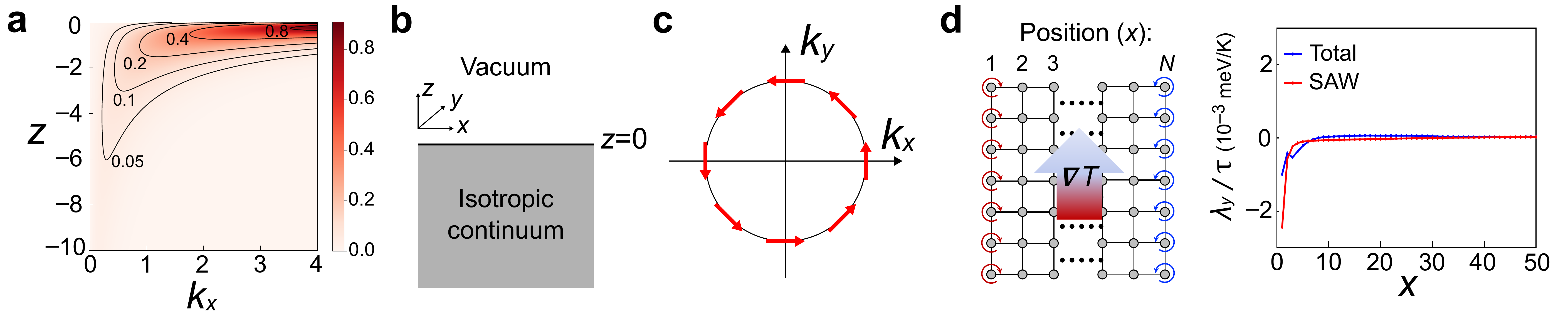}
\caption{\tbf{Phonon angular momentum of surface acoustic waves.}
\tbf{a,} Distribution of the angular momentum of the surface acoustic wave at given $k_x$ for the infinite slab geometry shown in \tbf{b}.
The color represents the density of angular momentum along the $y$ direction.
Because the sum over $z$ of the angular momentum along the $y$ direction for fixed $k_x$ is positive, we conclude that the angular momentum of the surface acoustic wave at a fixed energy is given as shown in \tbf{c}.
\tbf{d,} Left: square lattice model. The nearest neighbor longitudinal and transverse spring constants are $50$ and $20$, respectively.
The next nearest longitudinal and transverse spring constants are $50/4$ and $20/4$, respectively.
Right: By applying thermal gradient along the $y$ direction, there is an accumulation of phonon angular momentum at the edges.
For calculation, the total thickness was set to $300$.
We show the distribution near the left edge (the distribution of angular momentum is antisymmetric in the $x$ direction.)}
\label{Fig_S6}
\end{figure*}

\section{Phonon angular momentum at the surface}
Because both the phonon angular momentum Hall effect and the surface acoustic wave are properties of the low-energy phonons, it is tempting to ask whether there is any relation between them.
It is known that surface acoustic waves induce rotational motion~\cite{matsuo2013mechanical}, and we find that surface acoustic wave also have phonon angular momentum.
The reason behind this is that the inversion symmetry is broken whenever surface is introduced, as we explain below.
 
Let $\bu(\br)$ be the displacement and $\bp(\br)$ be the momentum of elastic medium at position $\br$.
For convenience, we choose to work with the rescaled position and momentum, $\bu(\br) \sqrt{\rho} \ra \bu(\br)$ and $\bp/\sqrt{\rho} \ra \bp(\br)$, where $\rho$ is the mass density.
We will always put $\hbar=1$.
In the momentum space, the low-energy phonon Hamiltonian (continuum approximation) is $\mc{H}^{(0)} = \frac{1}{2}\sum_{\bk} \bx_{\bk}^\dg H_{\bk} \bx_{\bk}$, where $\bx_{\bk} = \begin{psmallmatrix} \bp_\bk \\ \bu_\bk \end{psmallmatrix}$ and
\ba
H_\bk &= \bpm
1_{3\times 3} & 0 \\
0 & D_\bk
\epm.
\ea
Here, $(D_\bk)_{\alpha \beta}$ is the dynamical matrix.

The phonon angular momentum refers to the orbital motion of the atoms making up the solid with respect to the rest position.
For elastic continuum, we have $\bm{\mc{L}} = \int d\br \, \bu(\br) \times \bp(\br)$.
In the momentum space, $\mc{L}^\rho = \frac{1}{2} \sum_\bk \bx^\dg_\bk L^\rho \bx_\bk$, where
\ba
L^\rho=\bpm
0 & -\ell^\rho \\
\ell^\rho & 0
\epm, \quad
\ell^\rho_{\mu \nu} = \ep_{\mu \nu \rho}, \label{eq.pam_matrix}
\ea
and $\ep_{\mu \nu \rho}$ is the Levi-Civita symbol (we drop the zero-point phonon angular momentum as it is not our focus).
Since the time-reversal symmetry and inversion symmetry are represented by
\ba
\mc{T} = \bpm
-1_{3 \times 3} & 0 \\
0 & 1_{3 \times 3}
\epm \mc{K}, \quad \mc{P} = -1_{6 \times 6},
\ea
we see that the phonon angular momentum transforms like an axial vector in the momentum space.
Therefore, when both the time reversal and inversion symmetry are present, the phonon angular momentum vanish in the momentum space.

Let us now consider the surface acoustic wave in an isotropic medium propagating in the $x$ direction with surface termination at $z=0$.
Then, the mirror symmetry $\mc{M}_y:(x,y,z) \ra (x,-y,z)$ forces the angular momentum to point along the $y$ direction.
To calculate this component of the angular momentum, we normalize the polarization vectors such that $\int_{-\infty}^0 dz \, |\bep(k_x,z)|^2=1$.
Then the angular momentum density of the surface acoustic wave at $z$ in the $y$ direction is given by $L_{\rm SAW }(k_x,z)=-i \bep(k_x,z)^\dg \ell^y \bep(k_x,z)$.
For concreteness, let us assume that $v_T/v_L=0.5$, so that the surface acoustic wave has energy spectrum $E=k v_T \xi$ with $\xi \approx 0.933$, see \Sec{ssec.fbc}.
We show $L_{\rm SAW}(k_x,z)$ in \fig{Fig_S6}a.

Using the isotropicity of the phonon under consideration, the surface acoustic wave with $\bk= (k_x,k_y)$ therefore carries angular momentum $\bb L_{\rm SAW}(\bk,z)=L_{\rm SAW}(k,z) \hat{\bb z} \times \hat{\bk}$ as shown in \fig{Fig_S6}c.
The winding structure of the phonon angular momentum is reminiscent of the spin texture in Rashba electron gas, and we can expect that there is a thermal version of the Rashba-Edelstein effect \cite{bychkov1984properties,edelstein1990spin} for surface acoustic waves.
Note, however, that the bulk modes (non-localized) must also satisfy the boundary condition, so that it is not possible to neglect their contribution.

To see how much of the surface angular momentum results from SAW under the application of a thermal gradient, we compute the angular momentum induced by thermal gradient in a two-dimensional square lattice, as shown in \fig{Fig_S6}d (here, it suffices to consider only the in-plane vibration).
Here, we use the mass-spring model with the nearest neighbor longitudinal and transverse spring constants, and the next nearest neighbor longitudinal and transverse spring constants.
Let $D_\bk$ be the resulting $2N \times 2N$ dynamical matrix, where $N$ is the thickness, so that
\ba
H_\bk=\bpm
1_{2N \times 2N} & 0 \\ 
0 & D_\bk
\epm \label{eq.square_hamiltonian}.
\ea
It is also convenient to use the notation $|n,\bm{k}\rangle$ for phonon eigenstates, which satisfy
\ba
\sg^y H_\bk \ket{n,\bk} = E_{\bk,n} \ket{n,\bk}, \quad \bra{n,\bk} \sg^y \ket{m,\bk} = (\sg^z)_{nm}.
\ea
Here, $\sg^i$ with $i=x,y,z$ are the generalized Pauli matrices with block structure of $H_\bk$ in \eq{eq.square_hamiltonian}.
Note that there are $2N$ eigenstates with $n=-2N,-1,1,..., 2N$.

Let us apply a thermal gradient along the $y$ direction and compute the phonon angular momentum induced in the $z$ direction.
For a rough estimation of the thermally induced phonon angular momentum, we use the Boltzmann transport theory with constant phonon lifetime.
Then, the phonon angular momentum density induced by temperature gradient $(\nabla_y T) \hat{\bb y}$ is $\langle L^z_x \rangle_{\rm neq} - \langle L^z_x \rangle_{\rm eq}=-\lambda_y(x) \nabla_y T$, where
\ba
\lambda_y(x) =& \frac{\tau}{2k_BT^2}\frac{1}{V} \sum_\bk \sum_{n=-2N}^{2N} \bra{n,\bk} L_x^z \ket{n,\bk} \times \nn \\ 
& \bra{n,\bk} v_{\bk,y} \ket{n,\bk} \frac{\sg^z_{n,n} E_{\bk,n} e^{\sg^z_{nn} E_{\bk,n}/k_BT}}{(e^{\sg^z_{nn} E_{\bk,n}/k_BT}-1)^2}. 
\label{eq.pam_acc}
\ea
Here, $T$ is the temperature, $k_B$ is the Boltzmann constant.
$L^z$ is as given in \eq{eq.pam_matrix} with $\rho=z$ for each of the $N$ blocks (corresponding to each of the position $x$ in \fig{Fig_S6}), except that $\mu$ and $\nu$ takes only the values $x$ and $y$, since we are studying a two-dimensional model.

We show $\lambda_y(x)$ in \fig{Fig_S6}d. 
Because the lowest two energy bands (near $k_y=0$) are the surface acoustic waves for each of the two edges, we estimate the surface acoustic wave contribution by summing over the two lowest energy states ($n=1,2$) and the hole partners ($n=-1,-2$).
We see that the angular momentum due to the surface acoustic waves quickly decays, in contrast to that resulting from the bulk states, which does not decay to $0$ as fast.
We also see that the bulk states also contribute significantly to the surface angular momentum, although the trend of the total angular momentum induced on the surface is similar to that due to the SAW.

\bibliography{reference.bib}

\end{document}